\documentclass[twocolumn,pre,superscriptaddress]{revtex4-1}
\pdfoutput=1
\usepackage{amsfonts, amsmath, amsthm, amssymb}
\usepackage{xcolor}
\usepackage{soul}
\usepackage{hyperref}

\DeclareRobustCommand{\hltext}[1]{{\sethlcolor{yellow}\hl{#1}}}
\soulregister{\hltext}{1}

\usepackage{bbm}
\usepackage{stmaryrd}

\usepackage{graphicx}
\usepackage{float}
\graphicspath{{./figure/}}
\usepackage{graphics}
\usepackage{epstopdf}
 \usepackage{hyperref}

\usepackage{thmtools}
\usepackage{thm-restate}

\usepackage[]{appendix}

\usepackage{rotating}
\usepackage{fullpage}
\usepackage{natbib}  

\newcommand{\om}{\omega}

\newcommand{\beq}{\begin{equation}}
\newcommand{\eeq}{\end{equation}}
\newcommand{\barr}{\begin{array}}
\newcommand{\earr}{\end{array}}
\newcommand{\beqr}{\begin{eqnarray}}
\newcommand{\eeqr}{\end{eqnarray}}
\newcommand{\beqrn}{\begin{eqnarray*}}
\newcommand{\eeqrn}{\end{eqnarray*}}
\newcommand{\beqn}{\begin{equation*}}
\newcommand{\eeqn}{\end{equation*}}
\newcommand{\bei}{\begin{itemize}}
\newcommand{\beii}{\begin{itemize} \item}
\newcommand{\eei}{\end{itemize}}
\newcommand{\ben}{\begin{enumerate}}
\newcommand{\een}{\end{enumerate}}
\newcommand{\bes}{\begin{small}}
\newcommand{\ees}{\end{small}}
\newcommand{\bec}{\begin{center}}
\newcommand{\eec}{\end{center}}
\newcommand{\betab}{\begin{tabular}}
\newcommand{\eetab}{\end{tabular}}
\newtheorem{theorem}{Theorem}[section]
\newtheorem{lemma}[theorem]{Lemma}

\theoremstyle{definition}

\theoremstyle{remark}

\newcommand{\EV}{\mathbf{E}} 
\newcommand{\EVb}[1]{\EV\left\{#1\right\}} 
\newcommand{\EVs}[1]{\EV\left[#1\right]} 
\newcommand{\cov}{\mathrm{cov}}
\newcommand{\var}{\mathrm{var}}
\newcommand{\Var}{\mathbf{Var}}

\newcommand{\tr}{\mathrm{tr}}

\newcommand{\diag}{\mathrm{diag}}

\newcommand{\tnorm}[1]{\lVert#1\rVert_2}
\newcommand{\norm}[1]{\lVert#1\rVert}
\newcommand{\abs}[1]{\left\lvert#1\right\rvert}

\newcommand{\tsp}{\mathrm{T}}

\newcommand{\ER}{Erd\H{o}s-R\'{e}nyi}

\newcommand{\Nbc}{N^{BC}}
\newcommand{\Thbc}{\Theta^{BC}}
\newcommand{\kpbc}{\kappa^{BC}}
\newcommand{\mubc}{\mu^{BC}}
\newcommand{\Gbc}{G^{BC}}
\newcommand{\Ndeg}{N^{deg}}

\newcommand{\kpdeg}{\kappa^{deg}}
\newcommand{\mudeg}{\mu^{deg}}
\newcommand{\Gdeg}{G^{deg}}

\newcommand{\PF}{Perron-Frobenius}

\usepackage[framemethod=tikz]{mdframed}

\begin{document}
\title{Feedback through graph motifs relates structure and function in complex networks}

\author{Yu Hu}
\thanks{Current address: Dept. of Mathematics and Division of Life Science, the Hong Kong University of Science and Technology, Hong Kong, China}
\affiliation{Department of Applied Mathematics, University of Washington, Seattle, WA 98195}
\author{Steven L. Brunton}
\affiliation{Department of Applied Mathematics, University of Washington, Seattle, WA 98195}
\affiliation{Dept. of Mechanical Engineering, University of Washington, Seattle, WA 98195}
\author{Nicholas Cain}
\affiliation{Allen Institute for Brain Science, Seattle, WA 98109}
\author{Stefan Mihalas}
\affiliation{Allen Institute for Brain Science, Seattle, WA 98109}
\author{J. Nathan Kutz}
\affiliation{Department of Applied Mathematics, University of Washington, Seattle, WA 98195}
\affiliation{Dept. of Electrical Engineering, University of Washington, Seattle, WA 98195}
\affiliation{Dept. of Physics, University of Washington, Seattle, WA 98195}
\author{Eric Shea-Brown}
\affiliation{Department of Applied Mathematics, University of Washington, Seattle, WA 98195}
\affiliation{Dept. of Physiology and Biophysics, University of Washington, Seattle, WA 98195}
\affiliation{Allen Institute for Brain Science, Seattle, WA 98109}

\begin{abstract}
In physics, biology and engineering, network systems abound. 
How does the connectivity of a network system combine with the behavior of its individual components to determine its collective function?   
We approach this question for networks with linear time-invariant dynamics by relating internal network feedbacks to the statistical prevalence of connectivity motifs, a set of surprisingly simple and local statistics of connectivity. 
This results in a reduced order model of the network input--output dynamics in terms of motifs structures. 
As an example, the new formulation dramatically simplifies the classic \ER{} graph, reducing the overall network behavior to one proportional feedback wrapped around the dynamics of a single node. 
For general networks, higher-order motifs systematically provide further layers and types of feedback to regulate the network response.
Thus, the local connectivity shapes temporal and spectral processing by the network as a whole, and we show how this enables robust, yet tunable, functionality such as extending the time constant with which networks remember past signals. 
The theory also extends to networks composed from heterogeneous nodes with distinct dynamics and connectivity, and patterned input to (and readout from) subsets of nodes.
These statistical descriptions provide a powerful theoretical framework to understand the functionality of real-world network systems, as we illustrate with examples including the mouse brain connectome.
\end{abstract}

\maketitle


\section{Introduction}

Networked systems are ubiquitous across the physical, engineering, and biological sciences : including wave guide networks~\cite{Feigenbaum2010}, epidemic transmission~\cite{Castellano2010}, quantum networks~\cite{Rosset2016}, percolation and phase transitions~\cite{Gao2011}. These systems are characterized by a large connectivity graph that determines how the system operates as a whole~\cite{Watts:1998db,Park2013science,newman2003structure,albert2002statistical}.  
The connectivity is typically so complex that the structure-function relationship is obscured.  However, it is infeasible that every individual connection in a network is masterfully planned, or even necessary for functionality. Moreover, in many cases of practical interest it is impossible or exceedingly expensive to even completely measure the entire connectivity of a network. 
On the other hand, sampling a network via repeated, partial observations is often possible, and this has revealed an intriguing over-representation of certain types of localized connectivity patterns, or network {\it motifs}~\citep{Milo2002science,Song:2005jy,Perin:2011cu}. 
The alternative we explore here is that some statistical features of connectivity drive the underlying network function, motivating significant interest in studying specific connectivity patterns, or {\it network motifs}, that occur at higher than chance rates~\cite{Milo2002science,Song:2005jy}. 
The key insights from our theory are: (i) the {\it global} response of large complex networks to dynamic stimuli can  be predicted based on the statistics of  \emph{local} connectivity motifs, (ii) motifs of different sizes affect the network transfer function via distinct  temporal filters (Theorem \ref{TH:uniform_motif} and Fig. \ref{fig:ladder_diagram}B), and (iii) the effects from different motifs are combined nonlinearly but systematically to shape the overall network response  (Theorem \ref{TH:time_constant}).

This work draws new connections between two disciplines: (i) the statistical theory of networks ~\cite{Milo2002science,Alon:2007uu,Hu:2013vh} to isolate the impact of network motifs and (ii) control theory~\cite{Skogestad2005book} to describe the network response via an equivalent feedback circuit. There has been significant interest and effort designing robust distributed control of networked systems~\cite{Rahmani:SIAMJCO09,Newman10, Mesbahi10}, including multi-agent control for the internet~\cite{Low2002ieeecs,Doyle2005pnas} and the electric grid~\cite{Susuki2011jns}. However, there is relatively little work that relates network structure to function in the context of internal feedback and control theory~\cite{Liu2011nature}. Our result fills that gap, and shows how new network responses can be designed by tuning specific connection statistics.

To develop a concrete theory, we focus on the input-output properties of networks containing linear time-invariant (LTI) nodes. Such LTI networks are rooted in an extensive literature in control theory~\citep{Ogata:2010tz}, with a broad range of applications including: consensus and cooperation in networked multi-agent systems~\cite{Olfati2007procieee,Saboori2014ieeetac}, fault detection and isolation~\cite{Rahimian2015ieeetcns}, input localization~\cite{Nudell2015ieeetac}, optimal control~\cite{Imer2006automatica}, neuronal and regional circuits in the brain~\cite{Ganguli:2008tf,Goldman:2009ko,Chaudhuri2015}.
Further, such linear models have been used to describe nonlinear systems around a steady state or periodic orbit~\cite{Skogestad2005book,dp:book, Imer2006automatica,Nudell2015ieeetac}. A similar set of equations as Eq.~\eqref{E:node} -- and the same theory we develop here --  can also be applied to describe a very widely used set of linearly interacting point process models (the Hawkes process~\cite{Hawkes:1971vr,Hawkes:1971wh}) (see Appendix~\ref{sec:point_process}). These models have, in turn, been used to describe nonlinear systems with pulsatile interactions such as spiking neural networks~\citep{Pernice:2011fr,Trousdale:2012bj,Lindner:2005hd}.

\begin{figure}[h]
\begin{center}
\includegraphics[width=.35\textwidth]{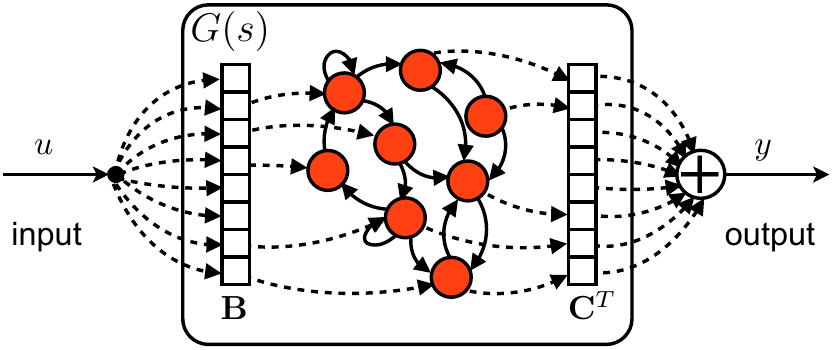}\\
\caption{A input signal $u(t)$ is sent to the network according to weights $B_i$, and a readout is formed by summing node activities with weights $C_j$.}
\label{fig:general_setup}
\end{center}
\end{figure}

As illustrated in Fig.~\ref{fig:general_setup}, we consider a network system consisting of $N$  nodes which are recurrently connected via a directed connectivity matrix $W$, whose entries can be real valued to represent graded connection weights. A scalar-valued, time-dependent input signal $u(t)$, is fed to the network according to a weight vector $B$. That is, each node $i$ receives an external input of $B_i u(t)$. We gather an output $y(t)$ by a linear combination of unit outputs $x_i(t)$ according to the vector $C$, so that $y(t)=\sum_{i} C_i x_i(t)$. The signal processing function of the network can then be characterized as the relationship between the input $u(t)$ and output $y(t)$. The dynamics of each LTI node is completely described by a temporal filter $h(t)$ and can be written as 

\beq
x_i(t)=\int_{0}^\infty h(\tau) \left(\sum_{j=1}^{N} W_{ij} x_j (t-\tau)+B_i u(t-\tau) d\tau \right).
\label{E:node}
\eeq

As a result of its having LTI nodes, it is easy to verify that the network as a whole is also a LTI system. In fact, we can derive  explicitly  the filter that the entire network applies to its inputs, hereafter denoted by $G(t)$. This is accomplished via the Laplace transform $\mathcal{L}(f)(s)=\int_{0}^\infty e^{-st} f(t) dt$ of Eq.~\eqref{E:node}, which allows us to rewrite the convolution conveniently as multiplication. Collecting outputs of nodes together as a vector $x(t)=(x_1(t),\ldots,x_N(t))^\tsp$, we have the following equation in matrix form:
\beq
x(s)=h(s)(Wx(s)+B u(s)).
\label{E:x_laplace_matrix}
\eeq
Here we overload the notation of $h(\cdot)$ with both the temporal filter $h(t)$ and the Laplace transform $h(s)$, and similarly for other variables and throughout the manuscript. The Laplace transform of a temporal filter offers an equivalent description of the LTI system in the frequency domain. Solving the system of Eq.~\eqref{E:x_laplace_matrix} gives the {\it network transfer function}
\beq
G(s):=\frac{y(s)}{u(s)}=C^\tsp (I-h(s)W)^{-1} B h(s).
\label{E:transfer_G_BC}
\eeq
Here $I$ is the identity matrix. Unless stated otherwise, we will consider the uniform input and output weights $B=C=(1,\ldots,1)^T/\sqrt{N}$.

\section{Network transfer function determined by motif cumulants}
\label{S:main_result}

\begin{figure}[h]
\begin{center}
\includegraphics[width=.48\textwidth]{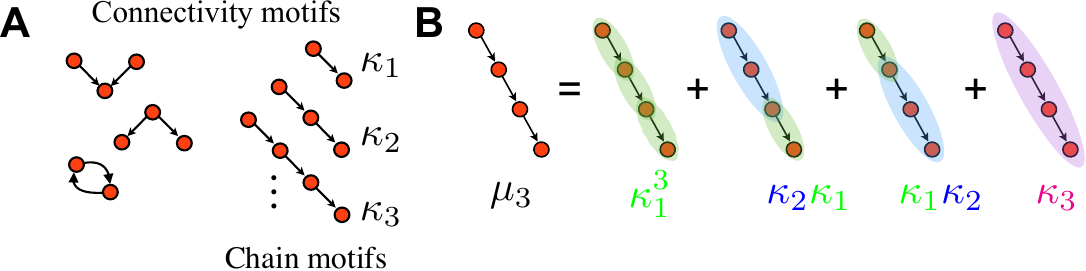}\\
\caption{ {\bf A}. Examples of connectivity motifs;  As we will show (for the basic case of Theorem \ref{TH:uniform_motif}), the network transfer function $G(s)$ is determined by the prevalence of chain motifs, given by the motif cumulants $\kappa_n$. {\bf B}. By decomposing motifs into smaller ones (shaded sub-motifs), the motif cumulants ($\kappa_3$ in this example) isolate the ``pure" higher order connectivity structure from raw motif counts ($\mu_3$ here).}
\label{fig:example_motifs}
\end{center}
\end{figure}

We now show how the connectivity $W$ determines the network transfer function $G(s)$. From Eq.~\eqref{E:transfer_G_BC}, it appears that all aspects of $W$, such as each entry or eigenvalue, may affect $G(s)$. 
However, we show that only a small, highly simplified set of statistical features of $W$ determine the network transfer function $G(s)$. 
These features are chain motifs, quantified via {\it motif cumulants}, a key tool introduced in ~\cite{Hu:2013vh, Hu:2014gs} to capture higher order connectivity structures in complex networks  (Fig.~\ref{fig:example_motifs}A). {\hl Motif cumulants, previously used to efficiently predict global levels of network synchrony from local connectivity structures  ~\cite{Hu:2013vh, Hu:2014gs} ,} quantify the frequency of ``pure" motif structures of a given size, over and above the frequency expected from smaller motifs that form its building blocks.  A positive motif cumulant indicates an over-representation of a certain motif, whereas a negative motif cumulant indicates an under-representation.  

Motif cumulants are closely related to simpler network statistics, the {\it motif moments}, which are defined by counting the number of occurrences of a motif in the network, and normalized by the number that would be in a complete (i.e., completely connected) graph. For example, the motif moment for length $n$ chains ($n$ consecutive connections among nodes $i_1\rightarrow i_2 \rightarrow \cdots \rightarrow i_n$) is $\mu_n=\sum_{i,j}(W^n)_{ij}/N^{n+1}$.  Following~\citep{Hu:2014gs}, the motif cumulant of $W$ for length $n$ chains, $\kappa_n$, can be (recursively) defined via the combinatorial decomposition relation
\beq
\mu_{n} =
\sum_{\{n_1,  \cdots , n_t\} \in \mathcal{C}(n)} \left(\prod_{i=1}^{t} \kappa_{n_i}\right).
\label{E:mu_kappa}
\eeq
Here $\mathcal{C}(n)$ is the set of all compositions (\emph{ordered} partitions) of $n$. An example of such a decomposition is shown in Fig.~\ref{fig:example_motifs}B.

Importantly, motif cumulants of order $n$, containing $n$ connections, can be estimated from sampling of the connectivity among up to $n+1$ nodes in the network (Appendix~\ref{S:estimating_motif_sampling}); thus, they are \emph{local} features of network connectivity, which can be key to quantifying them experimentally~\cite{Song:2005jy,Perin:2011cu}.

Besides the combinatorial definition, motif cumulants can also be calculated using a matrix expression~\citep{Hu:2014gs},
\beq
\kappa_n=\frac{1}{N^n}e^\tsp W (\Theta W)^{n-1} e, \quad\text{ where   } \Theta=I- e e^\tsp.
\label{E:kappa_matrix_ch}
\eeq

We are now ready to explain our main result that describes the relation between connectivity and the network transfer function $G(s)$ via the following theorem.

\begin{theorem}
\label{TH:uniform_motif}
A network transfer function $G(s)$ described by Eq.~\eqref{E:transfer_G_BC} with uniform input and output weights, that is, $B=C=(1,\ldots,1)^T/ \sqrt{N}$ can be written as
\beq
G(s)=\left(1-\sum_{n=1}^{\infty} N^n \kappa_n h^n(s)\right)^{-1} h(s),
\label{E:transfer_motifs}
\eeq
provided the connection strength is sufficiently small so that the series above converges (the condition for this being $\abs{h(s)} \rho(\Theta W\Theta)<1$, $\Theta=I-e e^\tsp$, $\rho(\cdot)$ is the spectral radius). Here $\kappa_n$ are chain motif cumulants defined in Eq.~\eqref{E:mu_kappa}.
\end{theorem}

\noindent We emphasize that Eq.~\eqref{E:transfer_motifs} is an exact expression that applies to any network connectivity $W$  (as long as the spectral radius constraint is satisfied), be this empirically measured or theoretically defined, and does not require any assumption of $W$ being sampled from certain ensembles of graphs. Moreover, Theorem~\ref{TH:uniform_motif} and the definition of motif cumulants Eq.~\eqref{E:mu_kappa} also apply to networks with non-uniform connection weights that vary from one link to another.  The motif cumulants are therefore interpreted as statistics for motifs, where each ``count"  of occurrence is weighted by the product of the strength of connections it contains.

The proof of Theorem \ref{TH:uniform_motif} is based on the combinatorial properties of $\kappa_n$ (Eq.~\eqref{E:mu_kappa}), similar to the approach used in~\citep{Hu:2014gs}.
\begin{proof}[Proof of Theorem~\ref{TH:uniform_motif}]
Starting from Eq.~\eqref{E:transfer_G_BC} with $B=C=e = (1,\ldots,1)^T/ \sqrt{N}$, we expand the matrix inverse as a power series
\[
G(s)=h\sum_{n=0}^{\infty}e^\tsp W^n e h^{n} =h\sum_{n=0}^{\infty} N^{n} h^{n} \mu_n.
\]
Substituting for $\mu_n$ with the decomposition Eq.~\eqref{E:mu_kappa} gives
\beqrn
G(s)&=&h\sum_{n=0}^{\infty} N^{n} h^{n} \sum_{\{n_1,  \cdots , n_t\} \in \mathcal{C}(n)} \left(\prod_{i=1}^{t} \kappa_{n_i}\right)\\
&=&h+h\sum_{n=1}^{\infty} \sum_{\{n_1,  \cdots , n_t\} \in \mathcal{C}(n)} \left(\prod_{i=1}^{t} (Nh)^{n_i}\kappa_{n_i}\right) \;.
\eeqrn
The summation above $\sum_{n=1}^{\infty} \sum_{\{n_1,  \cdots , n_t\} \in \mathcal{C}(n)}$ goes over all ordered partitions for each positive integer $n$ (exactly once). We now enumerate these ordered partitions in a different order: first, consider the number of components $t$ in the partition; next, note that each of these $t$ components can take any positive integer value $n_i$ ($i=1,\ldots, t$). Note further that each specified $t$ and $\{n_i\}$ corresponds to exactly one ordered partition in the original summation; and all ordered partitions will be enumerated for some $t$ and $\{n_i\}$. This shows that we can rewrite the summation in the following order

\beqrn
G(s)&=&h+h\sum_{t=1}^{\infty}\sum_{n_1,  \cdots , n_t =1}^{\infty} \left(\prod_{i=1}^{t} (Nh)^{n_i}\kappa_{n_i}\right).
\eeqrn
Note that the sum over $n_1,\cdots, n_t$ and the product summand can be factorized, which yields an identical factor for every $n_i$:
\beqrn
G(s)&=&h+h\sum_{t=1}^{\infty}\prod_{i=1}^{t}\left(\sum_{n_i=1}^\infty (Nh)^{n_i}\kappa_{n_i} \right)\\
&=&h+h\sum_{t=1}^{\infty}\left(\sum_{n=1}^\infty (Nh)^{n}\kappa_{n} \right)^t \; \;.
\eeqrn
Finally, summing the geometric series yields Eq.~\eqref{E:transfer_motifs},
\beqrn
G(s)&=&h+h\frac{\sum_{n=1}^\infty (Nh)^{n}\kappa_{n}}{1-\sum_{n=1}^\infty (Nh)^{n}\kappa_{n}}\\
&=&\frac{h}{1-\sum_{n=1}^\infty (Nh)^{n}\kappa_{n}}.
\eeqrn

\end{proof}

Theorem~\ref{TH:uniform_motif} provides a major simplification of the relationship between connectivity
$W$ and the network transfer function.  
First, only chain motifs appear in Eq.~\eqref{E:transfer_motifs}. This shows they are the only \emph{independent} connectivity features that affect $G(s)$. Other types of motifs and connectivity features may indirectly modify $G(s)$, but their effect is fully quantified in terms of their impact on chain motif cumulants.

Moreover, as we will illustrate below, the representation is highly efficient. Keeping only the first few terms of the infinite sum in Eq.~\eqref{E:transfer_motifs} can provide a good approximation of $G(s)$. This is in contrast to the slow convergence of a naive expansion of $G(s)$ in powers of $h(s)$, which would have coefficients related to motif counts $\mu_n$  instead of motif cumulants $\kappa_n$. Intuitively, the $\kappa_n$ decay rapidly with size $n$, as they have had any redundancy from their sub-components removed.  We empirically observe this fast decay in many graph models~\cite{Hu:2014gs}. This has important practical consequences, as the global network dynamics can then be explained in terms of a few measurable connection statistics.\\

\section{Networks reduced to first order ``motif" and proportional feedback}
\label{sec:proportional_feedback}
\begin{figure}[h]
\begin{center}
\includegraphics[width=.3\textwidth]{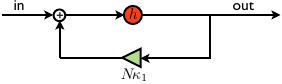}
\end{center}
\caption{The network transfer function for uniform degree networks and \ER{} networks are equivalent to proportional feedback around a single node.}
\label{fig:diagram_prop_feedback}
\end{figure}

To gain intuition for the formula in Theorem~\ref{TH:uniform_motif} and illustrate the powerful simplifications it provides, we first consider special, yet widely used, types of networks that can be reduced to only the ``trivial'' motif of first order chain cumulant $\kappa_1$ in the relation to the network transfer function.  For such networks,  the formula in Theorem~\ref{TH:uniform_motif} greatly simplifies, yielding a network transfer function $G(s)$ that  is precisely equivalent to proportional feedback on a single node.  This equivalent feedback diagram is shown in Fig.~\ref{fig:diagram_prop_feedback}.

\subsection{Small world, ``rotationally invariant," and other networks with uniform in- or out-degrees}
\label{sec:uniform_degree_network}
Networks with uniform in-degrees (that is, with the weighted sum of all incoming connections being the same for each node) or with uniform out-degrees are frequently described in neuroscience, physics, and in network science overall.  Such networks include regular networks \citep{Newman10}, and rewired regular networks.  Importantly, the latter is a popular model for small world networks, which have uniform out-degrees \citep{Watts:1998db}.  

Another way that networks with uniform degrees arise is through rotationally invariant connectivity structures. 
Rotationally invariant networks are characterized by defining a circular space variable $x$ and having the connectivity between two nodes depend on their spatial distance.  These occur in common models of, for example, neural networks that encode circular variables like direction of movement~\cite{Xie2002}.  Let's consider a network with rotationally invariant connection strengths $W_{ij}=w(j-i)$ and periodic boundary conditions $w(i)=w(i+N)$. The average connection weight is $\bar{w}=1/N \sum^{N}_{k=1} w(k)$.

Next, we  show that networks  with uniform in- or out-degrees have an interesting property:  their chain motif cumulants  $\kappa_n =0$ for all $n \ge 2$.   As a consequence, all of these networks produce input-output filters that are  equivalent to proportional feedback on a single node (Fig.~\ref{fig:diagram_prop_feedback} above). Intuition behind such a dramatic simplification comes from a combination of two facts. First, the dynamics of the nodes in the network is linear and thus the effect of potentially complex connectivity structures and pathways may be combined. Second, the fact that we send input to and read output from the network uniformly matches  the network's property of having uniform degrees. This allows the effect of connectivity be captured by its average, which is essentially the first term in Eq.~\ref{E:transfer_motifs} corresponding to the proportional feedback.

\begin{theorem}
For networks with uniform in- or out- degree, that is,
\[
\sum_{j=1}^{N}W_{ij}\equiv d_{in} \text{ or } \sum_{i=1}^{N}W_{ij}\equiv d_{out},
\]
all higher order chain motif cumulants are 0, that is, 
$
\kappa_{n\ge 2} =0.
$
\label{TH:uniform_deg_network_kappa_n}
\end{theorem}

\begin{proof}
Let's first consider a network with rotationally invariant connection strengths $W_{ij}=w(j-i)$ and periodic boundary conditions $w(i)=w(i+N)$. The average connection weight is $\bar{w}=1/N \sum^{N}_{k=1} w(k)$.
The second order chain moment is:
\begin{align}
\nonumber
	\mu_2 &=1/N^3 \sum_{i=1}^{N} \sum_{j=1}^{N} \sum_{k=1}^{N} W_{ij} W_{jk} \\
\nonumber	&=1/N^3 \sum_{i=1}^{N} \sum_{j=1}^{N} \sum_{k=1}^{N} w(j-i) w(k-j) \\
\nonumber	&=1/N^3 \sum_{i=1}^{N} \sum_{j=1}^{N} w(j-i) \sum_{k=1-j}^{N-j} w(k) \\ 
\nonumber	&=\bar{w}/N^2 \sum_{i=1}^{N} \sum_{j=1-i}^{N-i} w(j) \\
\nonumber	&=\bar{w}^2 \; .
\end{align}
Thus a rotationally invariant network has the same second order chain motif moment as a uniform network with $W_{ij}
\equiv \bar{w} \; \forall \, i,j$.

The $n$-th order ($n\ge 2$) chain moment is:
\begin{align}
\nonumber
	\mu_n &=1/N^{n+1} \sum_{i_1,i_n}^{N} \left(W^n\right)_{i_1,i_n} \\	
	&=1/N^n \sum_{i_1,i_{n-1},i_n}^{N} \left(W^{n-1}\right)_{i_1,i_{n-1}} W_{i_{n-1},i_n} 
	\label{end_chain}
	\\
	\nonumber
	&=1/N^{n} \sum_{i_1,i_{n-1}}^{N} \left(W^{n-1}\right)_{i_1,i_{n-1}} \sum_{i_n}^{N} W_{i_{n-1},i_n} \label{end_chain} \\ 
	\nonumber &=1/N^{n} \sum_{i_1,i_{n-1}}^{N} \left(W^{n-1}\right)_{i_1,i_{n-1}}  \bar{w}\\
	\nonumber  &=\mu_{n-1} \bar{w} \; .
\end{align}

Here we rewrite the chain motif $\mu_n$ in terms of a lower order motif $\bar{w} \mu_{n-1}$. Repeating this calculation, we conclude that $\mu_n  = \bar{w}^n$. Using the decomposition relation between $\mu_n$ and $\kappa_n$, we can show by induction that this indicates $\kappa_{n\ge2}=0$. Assume that $\kappa_{2\leq m \leq n-1} = 0$. All ordered partitions in Eq.~\eqref{E:mu_kappa} correspond to 0 except for when all $n_i$, $i=1,\ldots,n_t$ are either $1$ or $n$. This narrows down to two ordered partitions and  (note that $\kappa_1 = \bar{w}$)
\[
(\bar{w})^n = \mu_{n} = \prod_{i=1}^{n} \kappa_{1} + \kappa_n =  (\bar{w})^n + \kappa_n.
\]
This shows that $\kappa_{n} = 0$. By induction, we conclude all $\kappa_{n\ge2}=0$ in rotationally invariant networks.

The essential part of the proof was \eqref{end_chain}, in which the sum corresponding to the end of the chain was be factored out and summed, yielding the same value $N \bar w$ for each $i_{n-1}$.  This reduces the length of the chain by 1. In general, this step is possible as long as all the nodes in the network have the same (weighted) in-degree; this is the case of uniform in-degree.

Note that we can perform a similar reduction at the beginning of the product in \eqref{end_chain} instead of at the end, via
\[
1/N^{n+1} \sum_{i_1,i_n}^{N} \left(W^n\right)_{i_1,i_n}=1/N^n \sum_{i_1,i_{2},i_n}^{N}   W_{i_1,i_2} \left(W^{n-1}\right)_{i_2,i_{n}}. 
\]
Therefore, the same conclusion will follow if the network has uniform out-degree instead.

\end{proof}

\subsection{\ER{} graphs}
\ER{} (ER) random graphs~\citep{Erdos1961biis} are widely used in models of networked systems, where each connection is independently chosen to be present with probability $p$.  Interestingly, we show that for large \ER{} graphs, all non-trivial  motif cumulants vanish, just as for the uniform degree networks described in section~\ref{sec:uniform_degree_network}. 
\begin{theorem}
For an \ER{} graph with a fixed connection probability $p$, we have $\kappa_1\rightarrow p$ and $\kappa_{n\ge 2}\rightarrow 0$ in probability as 
$N\rightarrow \infty$.
\label{TH:ER_kappa}
\end{theorem}

\begin{proof}
By definition $\kappa_1=\frac{1}{N^2}\sum_{i,j}W_{ij}$. As the $W_{ij}$ are i.i.d. variables, $\kappa_1\rightarrow \EVs{W_{ij}}=p$ in probability as $N\rightarrow \infty $ by the law of large numbers.

For $n\ge 2$, using the matrix expression of motif cumulants Eq.~\eqref{E:kappa_matrix_ch}, 
\begin{eqnarray}
\nonumber
\abs{\kappa_{n}}&=&\abs{\frac{1}{N^n} e^\tsp (W\Theta)^{n-1} W e}\\
\nonumber
&\leq& \frac{1}{N^n} \tnorm{e^\tsp} (\tnorm{W\Theta})^{n-1} \tnorm{W} \tnorm{e} \\
\nonumber
&=& \left(\frac{\tnorm{W\Theta}}{N}\right)^{n-1} \frac{\tnorm{W}}{N}\\
&\leq& \left(\frac{\tnorm{W\Theta}}{N}\right)^{n-1} \frac{\norm{W}_{F}}{N}.
\label{E:kappa_n_bound}
\end{eqnarray}
Here $e=(1,\cdots,1)^\tsp/ \sqrt{N}$, and $\norm{W}_{F}=\sqrt{\sum_{i,j}W_{ij}^2}$ is the Frobenius norm. 

First, we show a bound on the second factor $\norm{W}_F/N$ in \eqref{E:kappa_n_bound}.
By the law of large numbers, for any positive $\delta$, 
\beq
\frac{1}{N^2}\sum_{i,j}W_{ij}^2 \leq \EVs{W_{ij}^2}+\delta=p+\delta
\label{E:W_Fnorm}
\eeq
 is satisfied with probability approaching 1 as $N\rightarrow \infty $; here, we used that $W_{ij}^2=W_{ij}$ since entries of the connection matrix are 0
 or 1. Choosing a fixed value of $\delta$ such as $\delta=p$, we have
 \[
 \norm{W}_F\leq \sqrt{2p}N,
 \]
 with probability approaching 1 as $N\rightarrow 1$.
 
 To finish the proof, we will use the following result to bound the first factor in \eqref{E:kappa_n_bound}. The proof of the lemma is given in Appendix~\ref{S:proof_ER_kappa}.
 
\begin{restatable}{lemmarestate}{thmWThtnorm}
For some absolute constant $C$, the probability that the following inequality holds approaches 1, as $N\rightarrow \infty$
\beq
\tnorm{W\Theta} \leq C\sqrt{N} \; \;.
\label{E:WTh_tnorm}
\eeq
\label{TH:WTh_tnorm}
\end{restatable}

Using Eq.~\eqref{E:W_Fnorm} and \eqref{E:WTh_tnorm} (along with a choice of the constant $\delta$), we see that the inequality 
\[
\abs{\kappa_n}\leq  C \left(\frac{1}{\sqrt{N}}\right)^{n-1} 
\]
holds with probability approaching 1, as $N\rightarrow \infty$, and for some  positive constant $C$ (independent of $n$ and $N$). Finally, as $C \left(\frac{1}{\sqrt{N}}\right)^{n-1} \rightarrow 0$ as $N\rightarrow \infty$, the above indicates that $\kappa_{n\ge2}\rightarrow 0$ in probability.
\end{proof}

We have seen that for networks with $\kappa_1$ as the only non-zero chain motif cumulant, the network transfer function can be represented as a proportional feedback. It turns out the ``reverse" is also true:  adding global feedback to an arbitrary network changes the network transfer function in the same way as adjusting its first motif cumulant.

Consider adding  a global feedback  by sending a proportion of the network output, $w_0 y(t)$, back to combine with its input $u(t)$.  One can directly verify that the new network transfer function for a network with such additional  feedback is 
\[ 
G^{\text{new}}(s) 
=\frac{h(s)}
{1-N(\kappa_1+  \frac{w_0}{N}) h(s) -\sum_{n=2}^{\infty} N^n \kappa_n h^n(s)}.
\]

This shows that the effect of such a global term, at the level of motif cumulants, is simply to shift $\kappa_1$ while keeping all $\kappa_{n\ge 2}$ the same. At the connectivity matrix level, the above modification in $\{\kappa_n \}$ corresponds to adding $w_0/N$ to all entries of $W$.

Furthermore, one can show that this equivalence holds not only in the sense of producing the same network transfer $G(s)$, but also in terms of the stability condition of the network, as explained in Appendix~\ref{sec:stability_mean_shift_W}). We will use these facts in Sec.~\ref{sec:stronger_k2_effect} to emphasize effects from higher order $\kappa_n$ by reducing $\kappa_1$.

\section{Networks with impact from higher order motifs $\kappa_{n\ge 2}$}
\label{sec:change_kappa2}

\begin{figure}[h]
\includegraphics[width=0.48\textwidth]{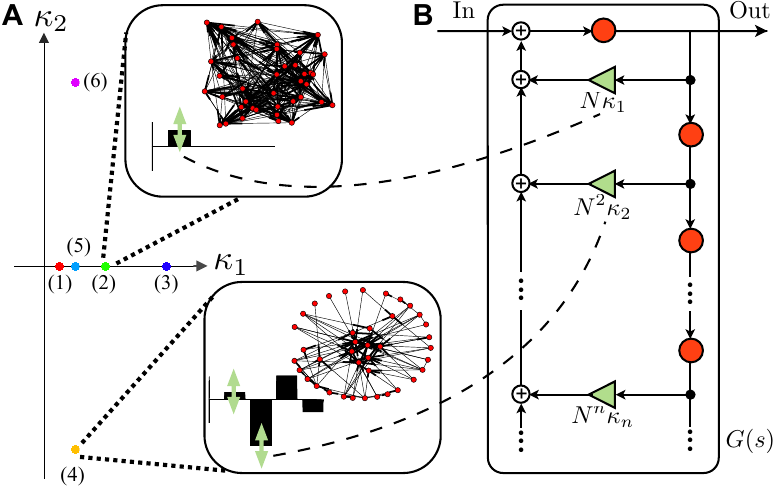} 
\caption{{\bf A}. Complex networks may be organized by their motif cumulants $\kappa_j$; the first two cumulants are shown.  We show two example networks with different motifs.  Bar graphs show values of $\kappa_1$ through $\kappa_4$ (relative magnitude to powers of the connection probability, see details in Appendix \ref{sec:motif_magnitude_scaling}),  for each network.  {\bf B} .  The motif content determines the strength of each pathway in the feedback hierarchy shown; this relationship is indicated by the green ``slider" arrows.
}
\label{fig:ladder_diagram}
\end{figure}

For complex networks where there are motif structures beyond first order ($\kappa_1$), the equivalence of connectivity to  feedback loops in Fig.~\ref{fig:diagram_prop_feedback} can be generalized: each motif cumulant gives rise to a unique feedback pathway, which combines to yield the ladder-structured control diagram shown in Fig.~\ref{fig:ladder_diagram}.  We emphasize that our usage of {\it motif cumulants} is essential:  by removing redundancy due to shorter component paths, each motif cumulant corresponds to a unique feedback link, instead of appearing at multiple links.

The functional diagrammatic representation in Fig.~\ref{fig:ladder_diagram} suggests that motif cumulants can be thought of as \emph{feedback knobs}, which shape the input-output properties of a network. In general, the impact of any motif cumulant depends on the presence or absence of others in a given network.  Moreover, motifs of different sizes affect the network transfer function in different ways. This is because the feedback link for $\kappa_n$ involves passing through $n$ copies of the node filter (the $h^n(s)$ factors in Eq. \eqref{E:transfer_motifs}).
 
\begin{figure}[h]
\begin{center}
\includegraphics[width=.45\textwidth]{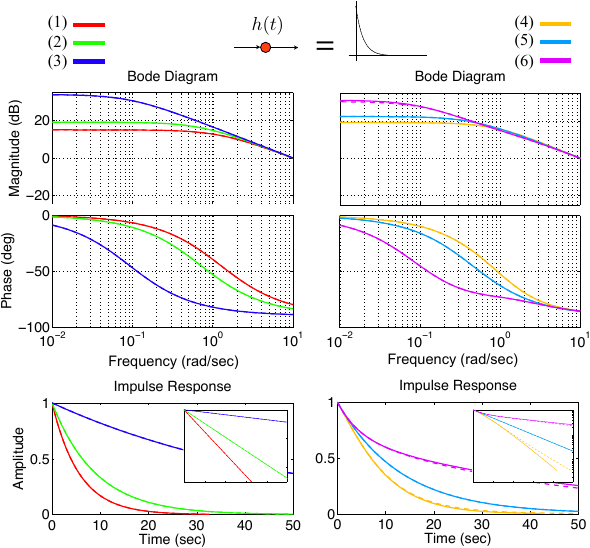}
\end{center}
\caption{ Network transfer functions $G(s)$ for different networks, indicated by matching color/shade and numbering as the dots in Fig.~\ref{fig:ladder_diagram} (see legends on top of each column); in the first column, networks have differing values of $\kappa_1$; in the second, differing values of $\kappa_2$.  Here, the node filter is $h_{\exp}$. Dashed lines are approximations by keeping leading terms (1 term in the left column and 3 terms for the right, see text); some are indistinguishable from the solid corresponding to actual filters with all terms.}
\label{fig:motif_effect_hexp}
\end{figure}

\begin{figure}[h]
\begin{center}
\includegraphics[width=.45\textwidth]{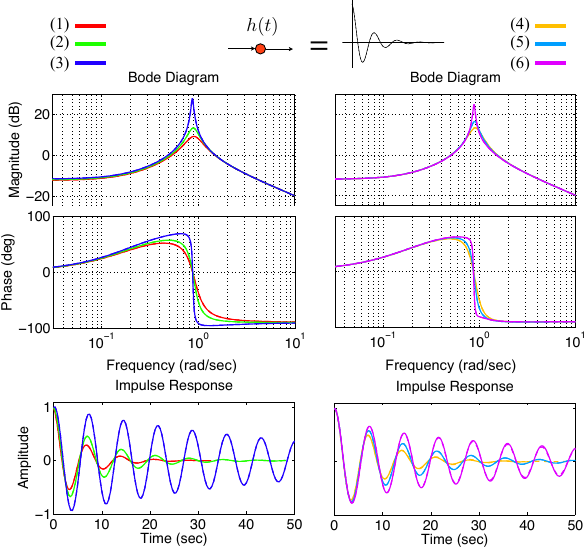}
\end{center}
\caption{Same as Fig.~\ref{fig:motif_effect_hexp} except that the node filter is $h_{\cos}$.}
\label{fig:motif_effect_hcos}
\end{figure}

To demonstrate this, we generate ``binary" networks where all non-zero connections have the same strength $a$, and their motif statistics lie in different locations on the plane of $\kappa_1$ and $\kappa_2$ (Fig.~\ref{fig:ladder_diagram}A, the networks are generated as ER networks or \emph{second order networks} (SONETS~\cite{Zhao:2011dv}), see Appendix~\ref{S:network_generation} for more details).
We fix all other parameters such as the coupling strength $a$ so that the only difference is the graphical structure of $W$. 
For concreteness, we set the node filter $h(t)$
to be either  
an exponential filter $h_{\exp}(t) = e^{-t/5}, t\ge 0$ (Fig.~\ref{fig:motif_effect_hexp}), or a decaying-oscillatory filter $h_{\cos}(t)$ (Fig.~\ref{fig:motif_effect_hcos})

Fig.~\ref{fig:motif_effect_hexp} and \ref{fig:motif_effect_hcos} show the change in the network transfer function for various $\kappa_1$ and $\kappa_2$ in both frequency and time domains. In the frequency domain, we use the standard Bode diagrams which plot the magnitude and phase of $G(s)$ along frequencies $s=i\omega$~\cite{Ogata:2010tz} (first two rows in Fig.~\ref{fig:motif_effect_hexp} and \ref{fig:motif_effect_hcos}); In the time domain, we plot the impulse response, that is the network's output given a brief impulse, which is also the inverse Laplace transform of $G(s)$, $G(t) =\frac{1}{2\pi i}\int_{-i\infty}^{+i\infty}   G(s)ds$   (bottom row in Fig.~\ref{fig:motif_effect_hexp} and \ref{fig:motif_effect_hcos}).
Increasing $\kappa_1$ while $\kappa_{n\ge2}\approx 0$, or equivalently increasing the connection probability in a ER graph, we observe a change in the network transfer function $G(s)$ from a low-pass filter towards an integrator (i.e. $G(s)=1/s$) in the case of $h(t)=h_{\exp}(t)$, and an increase of the magnitude of resonant peak in the case of $h(t)=h_{\cos}(t)$.
In the time domain, the impulse response correspondingly has a slower decay, in the case of $h_{\exp}(t)$, indicating an increased ``memory" to past inputs (a point we will return to), and enhanced oscillations in the case of $h_{\cos}(t)$.

Next, we change the connectivity $W$ along the $\kappa_2$ direction, while fixing $\kappa_1$. This is equivalent to changing the frequency of two-link chain motifs, while keeping the number of connections the same. Fig.~\ref{fig:motif_effect_hexp} and \ref{fig:motif_effect_hcos} show that this structural change achieves similar input-output dynamics as adding more connections to a ER graph.  Moreover, including a higher order motif cumulant, $\kappa_2$, can introduce new effects in $G(s)$ not present with $\kappa_1$ alone.  
For example, an enhanced frequency of two-link chains (positive $\kappa_2$) in networks with $h_{\exp}(t)$ nodes introduces additional timescales in the network impulse response, which is no longer described by a single exponential (Fig.~\ref{fig:motif_effect_hexp} third row insets on log magnitude plots).

More dramatic effects of $\kappa_2$ can be achieved for dense weighted networks, or for sparse networks via a modified version of the theory. We explain these networks and associated theory in Sec.~\ref{S:degree_corrected_motif}.

To confirm the efficiency and accuracy of our motif cumulant approach, we approximate $G(s)$ by truncating Eq. \eqref{E:transfer_motifs} to include only cumulants of up to size 3. Despite our using only highly local connectivity information, the resulting $G(s)$ (dashed lines in Fig.~\ref{fig:motif_effect_hexp}) functions closely match the exact $G(s)$ from the full connectivity matrix $W$ (solid lines).

Interestingly, many of these effects by changing motifs may be understood analytically via the movement of poles in the complex plane, a classic tool from control theory. The method links the properties of a transfer function $h(s)$ to its poles: the complex values of $s$ where the denominator of $h(s)$ becomes zero. Based on Eq.~\eqref{E:transfer_motifs}, the poles of $G(s)$ is closely related to the roots of a $m$-degree polynomial having motif cumulants as coefficients $P(z) = N^m \kappa_m z^m+ 
N^{m-1}\kappa_{m-1} z^{m-1}+\cdots +N\kappa_1 -1$, if $\kappa_{n>m} \approx 0$ and we can neglect the effect of those higher order motifs. A detailed analysis based on this (Appendix~\ref{S:pole_kappa}) explains the change in the speed of temporal decay and the amount of oscillations see in Fig.~\ref{fig:motif_effect_hexp} and \ref{fig:motif_effect_hcos}, as well as the ``bifurcation" occur with negative $\kappa_2$ for those networks with stronger effects from motifs in Sec.~\ref{S:degree_corrected_motif}.

\section{Motif cumulants and the time constant of system response}
\label{S:time_constant}

One intensely studied property of network systems in the literature~\citep{Ganguli:2008tf,Goldman:2009ko,Chaudhuri2015} is the network time constant: how long the network maintains a ``memory" of past signals once they have been removed. We will in particular study how connectivity statistics change the time constant in the context of mesoscale mouse brain network in Sec.~\ref{sec:real_networks}. We first develop a general theory that establishes a direct and explicit form relation between motif cumulants and the time constant. 

We quantify the timescale for a general filter filter via the ``frequency-cutoff" time constant. Specifically, in the Bode magnitude plot (which is in logarithm scale for both coordinates, e.g. Fig.~\ref{fig:motif_effect_hexp}), we draw a horizontal line at the level of ``baseline" gain (i.e., the magnitude at 0 frequency), and another asymptotic line following the decay at high frequencies; the x-coordinate of the intersection is the cut-off frequency $s_0$ (or its logrithm $\log_{10}(s_0)$). Intuitively, this is where the transition between a sustained response vs. a strongly damped response occurs. The time constant can in turn be defined as the reciprocal of the cut-off frequency $\tau = s_0^{-1}$.

This definition of time constant is consistent with the notion of the speed of temporal decay. Taking the exponential filter $h_{\exp}(t)=e^{-\alpha t },\; t\ge 0$ as an example, it is easy to verify that its time constant defined by the cut-off is $1/\alpha$, the same as the usual definition for the time constant of an exponential decay. 

The cut-off time constant for the network response $G(t)$ can be precisely linked to motif cumulants using the resumming formula Eq.~\eqref{E:transfer_motifs}.
\begin{theorem}
\label{TH:time_constant}
Consider a network with a node filter $h(s)$ that decreases asymptotically as $1/s^g$ ($g>0$) for large $s$, with a time constant $\tau_h$. Then the time constant of the network transfer function $G(s)$ is 
\beq
\tau_G = \left[ \tau_h^g / \left({1-N\kappa_1 \tau_h^g-N^2\kappa_2 \tau_h^{2g}-\cdots}\right) \right]^{1/g}.
\label{E:tau_motifs}
\eeq
\end{theorem}

\noindent We note that one  insight from Eq. \eqref{E:tau_motifs} is that the contributions from different motif cumulants are combined \emph{nonlinearly} to determine the network time constant. This is due to the appearance of $\kappa_n$ in the denominator. For example, the effect of changing $\kappa_2$ on the time constant will also depend on the value of $\kappa_1$.

\begin{proof}
First, we express the time constant of $h(s)$ according to the definition given above.
Because $h(s)\approx \frac{1}{s^{g}}$ for large $s$, the large frequency asymptotic line in the Bode plot of $h(s)$ is  $y=-20 g x$, where $x=\log_{10}s$ (note that the y-axis unit is decibels, hence the coefficient $20$). The low frequency asymptote of $h(s)$ is a horizontal line $y=20\log_{10} h(0)$. We can solve for the intersection of the two asymptotes:  they intersect on the Bode plot at $x=-g^{-1} \log_{10} h(0)$, corresponding to a cuttoff frequency $s_0=(h(0))^{-\frac{1}{g}}$. The time constant, which is the reciprocal of $s_0$, is $\tau_h=(h(0))^{\frac{1}{g}}$.

Now, we apply the similar calculation of intersection to $G(s)$. The baseline gain $G(0)$ can be calculated by setting $s=0$ in Eq.~\eqref{E:transfer_motifs} and substitute $h(0)$ with $\tau_h^g$ using the relation derived above. This gives $G(0) = \tau_h^g / \left({1-N\kappa_1 \tau_h^g-N^2\kappa_2 \tau_h^{2g}-\cdots}\right) $. And for large $s$, $G(s)\approx \frac{1}{s^{g}}$. Combining these, the time constant of $G(s)$ is
\[
(G(0))^{\frac{1}{g}}=\left[ \tau_h^g / \left({1-N\kappa_1 \tau_h^g-N^2\kappa_2 \tau_h^{2g}-\cdots}\right) \right]^{1/g}.
\]
\end{proof}

\begin{figure}[h]
\begin{center}
\includegraphics[width=0.48\textwidth]{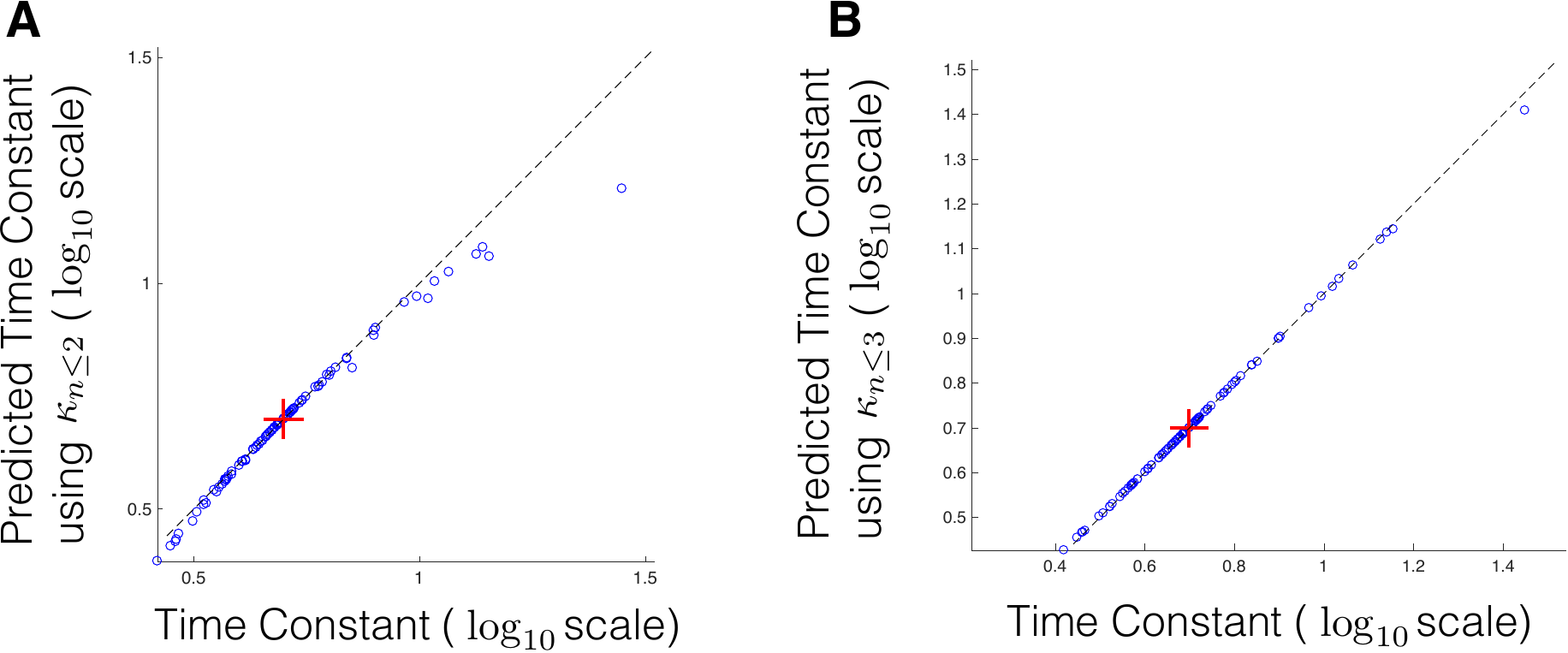}
\caption{Comparison of numerical (x-axis) and analytical predictions (y-axis) of time constants in 100 SONETs (each network sample is a dot). The motif based prediction is from truncating after second order {\bf A}) or third order ({\bf B}) motif cumulants, using Theorem~\ref{TH:time_constant}. The axes are in $\log_{10}$ scale, and the values are normalized by the time constant of a single node (which corresponds to 0). All the networks have the same connection strength and approximately the same number of connections (connection probability 0.1) by construction, but have various extents of second order motif cumulants. For comparison, the time constant for an \ER{} network is labeled by a red plus sign.
} 
\label{fig:time_constant_scatter}
\end{center}
\end{figure}

To test the utility of Theorem~\ref{TH:time_constant}, we numerically computed the time constant (by definition) for a large set of \emph{second order networks} (SONETs \cite{Zhao:2011dv}) with diverse values of $\kappa_2$, while $\kappa_1$ and the coupling strength of each connection are set to be the same for all networks. When compared with the approximations computed using Theorem~\ref{TH:time_constant} and only using $\kappa_1$ and $\kappa_2$, with higher order terms truncated, we see a very good agreement (Fig.~\ref{fig:time_constant_scatter}A). The motif based approximation can be further improved by keeping more cumulant terms in Theorem~\ref{TH:time_constant} (Fig.~\ref{fig:time_constant_scatter}B). We also observe a broad range of time constants spanning several orders of magnitude. As a comparison, the time constant for a matching \ER{} graph (with the same number of connections) is marked in Fig.~\ref{fig:time_constant_scatter}, and this would  also be the prediction based on $\kappa_1$ alone. This shows motif cumulants at various orders can have large impacts on timescale.  

\section{Network transfer in the presence of heterogeneous, random input and output weights} 

\subsection{Independent, random weights and the robustness of $G(s)$ in large networks}
\label{S:rand_weights}
Until now we have considered the case where the input and output weights are uniform over all the nodes in the network. Here we consider whether the results for these uniform weights are robust to noise in the weights. We start by considering the case where input and output weights $B_i$, $C_j$, for all $i,j=1,\cdots,N$ are independent and identically distributed (i.i.d.) variables 

For an arbitrary set of weights $B,C$, the matrix formula of $G(s)$, Eq.~\eqref{E:transfer_G_BC}, still applies. However, the expression Eq.~\eqref{E:transfer_motifs}, in terms of motif cumulants, no longer holds directly. Nonetheless, we can verify easily that the expectation $\EVs{G(s)}$ is essentially the same as the case with uniform weights and hence Eq.~\eqref{E:transfer_motifs} applies.
\beq
\EVs{G(s)} = \EVs{B_i} \EVs{C_j} \cdot e^\tsp (I-h(s)W)^{-1} e h(s),
 \label{E:E_Gs_ind}
\eeq
where $ e =(1,\ldots, 1)^T / \sqrt{N}$.

\begin{figure}[h]
\begin{center}
\includegraphics[width=0.3\textwidth]{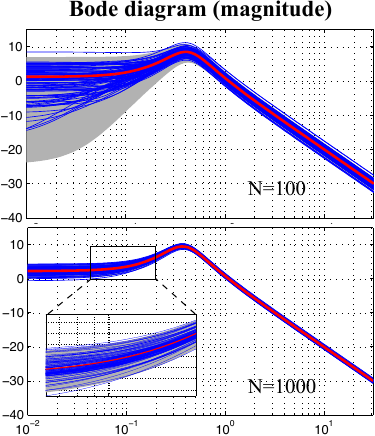}
\caption{Bode diagram showing convergence of network transfer functions in large networks; Blue thin lines are 100 individual networks. The red thick lines in the middle correspond to uniform input and output weights. The shaded area shows the 90\% confidence interval (Eq.~\eqref{eq:gaussian_CI_bound}). }
\label{fig:random_weights_convergence}
\end{center}
\end{figure}

Interestingly, for large network, we prove the following result that the random $G(s)$ converges to its expectation (proof given in Appendix~\ref{sec:proof_converge_G}), so the motifs cumulant description Eq.~\eqref{E:transfer_motifs} also describe each random $G(s)$ closely (Fig.~\ref{fig:random_weights_convergence}).

\begin{restatable}{thmrestate}{thmconverge}
\label{Th:converge}
Let $\kappa_n$ be the motif cumulants of a sequence of $W$, whose size $N\rightarrow \infty$. Assume that each $\kappa_n$ has a limit $\kappa_n^\infty$ as $N\rightarrow \infty$. Additionally, we assume a bound on the norm of $W$,
 \beq
 \tnorm{W}\leq (1-\delta)\frac{ N}{\max_{s}{\abs{h(s)}}}, 
\label{E:W_norm_condition}
 \eeq
for some fixed positive constant $\delta$. Let $G(s)$ be the (random) transfer function for networks with connection matrix $\frac{1}{N} W$, and random i.i.d. input/output weights $B, C$ with mean $\theta=\frac{1}{\sqrt{N}}$ and variance $\sigma^2=\frac{\sigma_0^2}{N}$ ($\sigma_0$ is a constant).  Then, we have the following convergence network transfer function as $N\rightarrow \infty$:
\beq
G(s)\rightarrow  G^\infty(s) \quad \text{ uniformly in $s$},
\label{E:G_limit}
\eeq
where $G^\infty(s)=\left(1-\sum_{n=1}^{\infty} h^n(s)\kappa^\infty_n \right)^{-1}h(s)$.
\end{restatable}

In addition to the asymptotic convergence, it is also important for applications to assess the rate of this convergence for a finite size network. To this purpose, we derive an estimate of the confidence interval describing the fluctuations of $|G(s)|$.

Our calculation is based on the ansatz or assumption that $G(s)$ for each $s$ is Gaussian distributed. The assumption is intuitively justified as $G(s)$ is a linear sum of a large number of random variables, $B_i, C_j$, when $W$ and $s$ are fixed. This assumption also appears to holds well in our numerical simulations.

\begin{figure}[h]
\begin{center}
\includegraphics[width=0.3\textwidth]{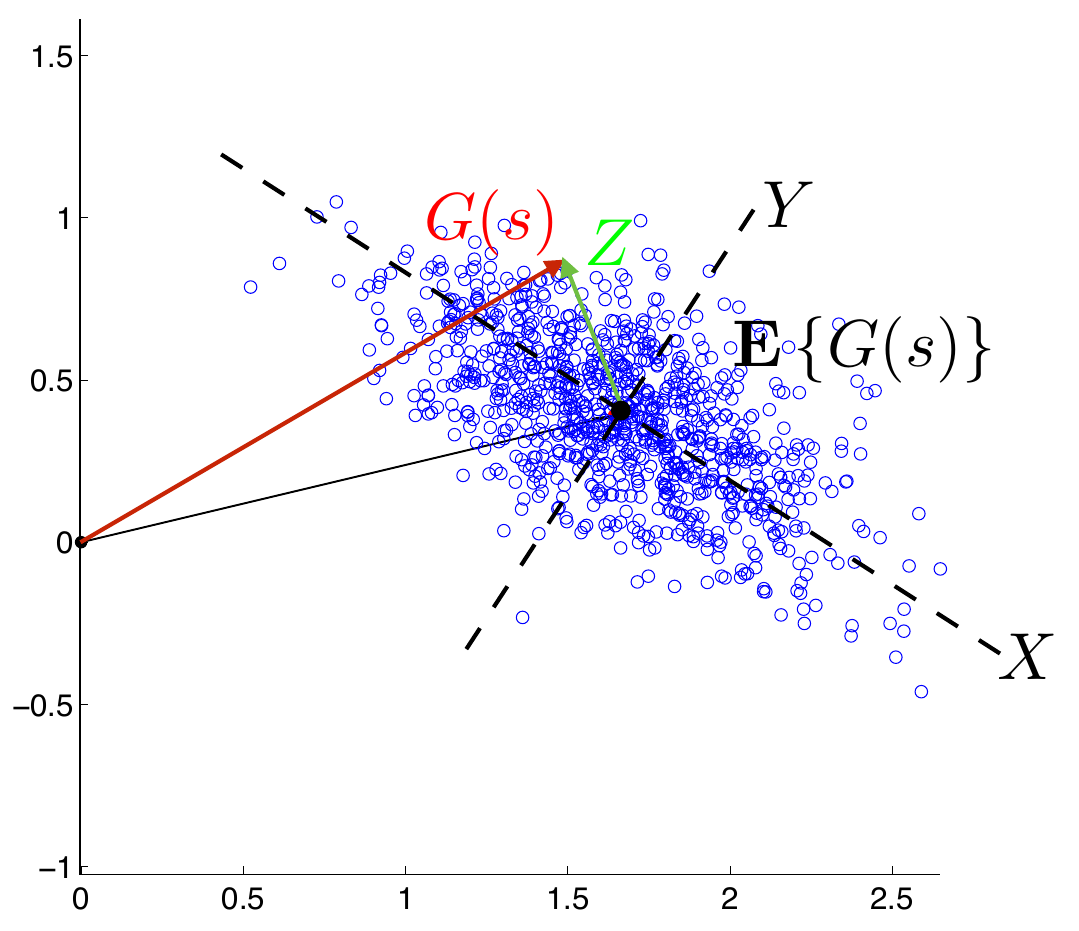}\\
\caption{Schematic of the distribution of $G(s)$ around $\EVs{G(s)}$ for a fixed $s$ and the related decomposition into independent components.}  
\label{F:G_s_dist}
\end{center}
\end{figure}

Note that the $G(s)$ in general are complex variables. Therefore, $G(s)$ will be Gaussian distributed in 2D under our assumption (Fig.~\ref{F:G_s_dist}). As $G(s)$ can have correlated components which may not be aligned with $\EVs{G(s)}$ (see Fig.~\ref{F:G_s_dist}), calculating the exact confidence interval that $G(s)$ lies in for a certain probability is quite involved. Instead we derive a simple upper bound.  

First, the triangle inequality gives
\begin{align}
\nonumber
&\abs{\EVs{G(s)}}-\abs{G(s)-\EVs{G(s)}} \leq \abs{G(s)}\\
&\leq \abs{\EVs{G(s)}}+\abs{G(s)-\EVs{G(s)}}.
\label{E:triangular_G}
\end{align}
Let $Z=G(s)-\EVs{G(s)}$ be the 2D random vector, which can be decomposed into independent real Gaussian components $X,Y$ such that $\abs{Z}^2=X^2+Y^2$. For $0<p<1$, let $\alpha=\Phi^{-1}(1-p/2)>0$, where $\Phi^{-1}(x)$ is the inverse cumulative distribution function for a standard Gaussian variable. Therefore, we have
\[
P(\abs{X}\leq \alpha \sigma_X)=1-p, \quad P(\abs{Y}\leq \alpha \sigma_Y)=1-p.
\]
Here $\sigma_X^2$ and $\sigma_Y^2$ are the variances of $X,Y$. Since $X$ and $Y$ are independent, 
\[
P(\abs{X}\leq \alpha \sigma_X,\abs{Y}\leq \alpha \sigma_Y)\leq (1-p)^2.
\]
Under this event, 
\[
\abs{Z}=\sqrt{X^2+Y^2}\leq \alpha \sqrt{\sigma_X^2+\sigma_Y^2}=\alpha \sigma_{Z}.
\]

Finally, using inequality Eq.~\eqref{E:triangular_G}, we arrive at the following upper bound of the confidence interval for $\abs{G(s)}$
\begin{eqnarray}
\nonumber
&&P\left(\abs{\EVs{G(s)}}  -   \alpha\sigma_{Z}   \leq \abs{G(s)}  \leq  \abs{\EVs{G(s)}}  +   \alpha\sigma_{Z}  \right)\\
&&\ge  (1-p)^2 
\label{eq:gaussian_CI_bound}
\end{eqnarray}
Here $\alpha=\Phi^{-1}(1-p/2)$, and $\sigma_{Z}^2$ is the variance of $Z=G(s)-\EVs{G(s)}$ whose expression is explicitly given in Eq.~\eqref{E:var_rand_ind_weight} in Appendix~\ref{sec:proof_converge_G}.
As shown in Fig.~\ref{fig:random_weights_convergence}, this estimate of the confidence interval indeed agrees well with numerical simulations.

\subsection{Correlated, random input and output weights recruit cycle motifs that determine network transfer}

We next consider the case when the input and output weight vectors $B,C$ are random but {\it correlated}.
Specifically, we take $B_i$ and $C_i$ to be correlated for each $i=1,\cdots,N$ while $B_i$ and $C_j$ are still independent for $i \neq j$. Such a correlation structure can be motivated in neuroscience, where plasticity mechanisms may lead more-active cells to both receive stronger inputs and more strongly influence cells downstream. 

\begin{figure}[h]
\begin{center}
\includegraphics[width=0.27\textwidth]{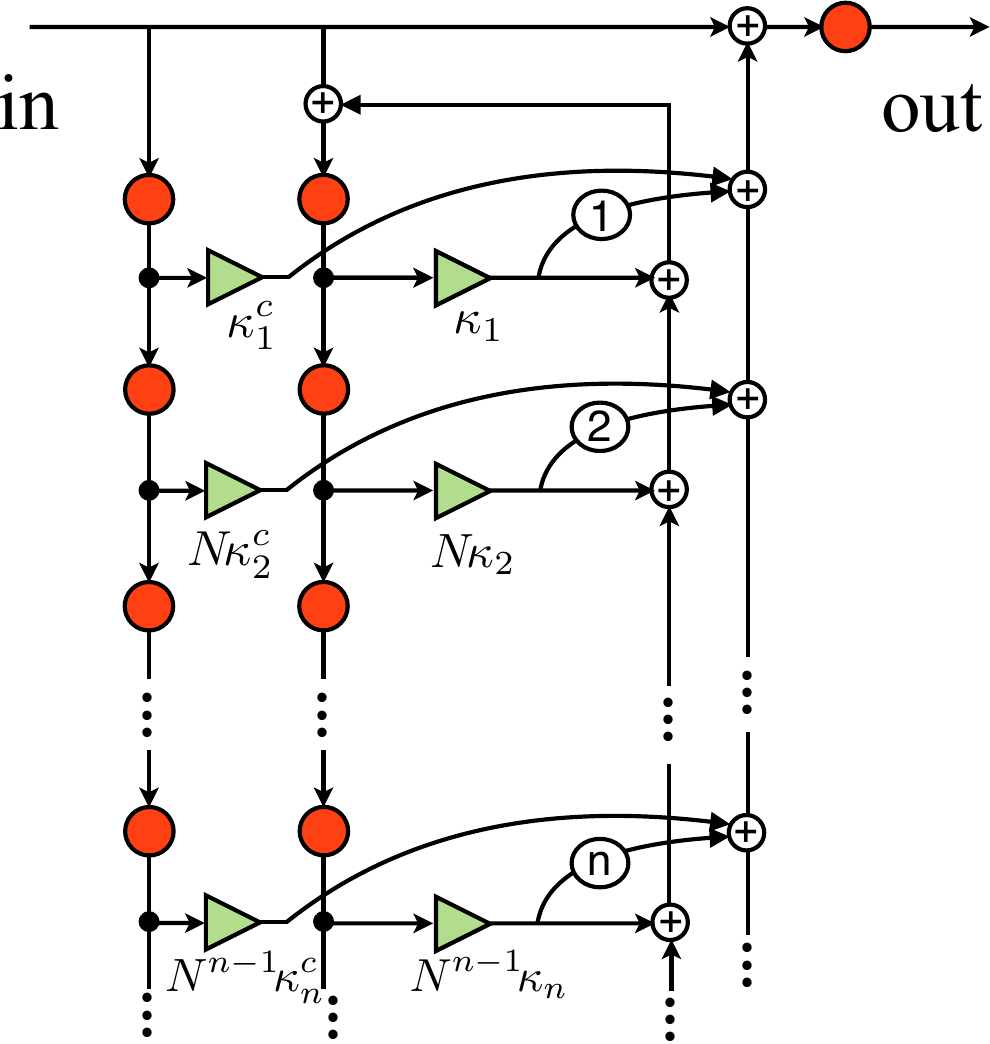}
\caption{Feedback diagram for a network with correlated input and output weight vectors $B,C$ (see text). Both chain and cycle motif cumulant gives rise to a feedback link in the flow diagram. The circled numbers are additional constant weights for some feedback links.}
\label{fig:cycle_diagram}
\end{center}
\end{figure}

As we describe below, the presence of correlations between input and output weights changes the average network transfer function $\EVb{G(s)}$, but it can still be described by a similar motif cumulant expression (Eq.~\eqref{E:trace_resum}), now involving additionally the \emph{cycle motif cumulants}. Moreover, the expression again has a feedback diagram interpretation (Fig.~\ref{fig:cycle_diagram}).

We first derive the expression for $\EVb{G(s)}$ in the correlated case. Let $\var(B_i)=\var(C_i)=\sigma^2$ and $\rho=\cov(B_i,C_i)/ \sigma^2$ be the correlation coefficient that is same for all $i=1,\cdots, N$. 
The following identity holds for any matrix $M$:
\begin{align*}
&\EVs{ B^T M C}  = \sum_{i,j}M_{ij} \EVs{  B_i C_j} \\
  &=  \sum_{i,j}M_{ij} (\theta^2 + \rho \sigma^2 \delta_{ij})
  = N \theta^2 e^T M e + \rho \sigma^2 \tr(M).
\end{align*}
Here $\tr(\cdot)$ is the trace and $e=(1,\ldots,1)^T/\sqrt{N}$. Using this identity
\begin{align}
\nonumber
&\EVs{G(s)}
=N\theta^2 e^\tsp (I-h(s)W)^{-1} e h(s)\\
&+\rho \sigma^2 \tr((I-h(s)W)^{-1})h(s).
\label{E:E_Gs_corr}
\end{align}
The appearance of the second term above reflects the correlation between input and output weights. If $\theta$ and $\sigma$ have the  same scaling with respect to $N$, the first term in Eq.~\eqref{E:E_Gs_corr} will dominate for large $N$. 
However, if the weights are balanced with both positive and negative values, giving a mean weight $\theta=0$, $\EVs{G(s)}$ will contain only the second term $\rho \sigma^2 \tr((I-h(s)W)^{-1})$. To focus on the effect of this new term, we set $\theta=0$ from now on unless stated otherwise.

Using the matrix Taylor series expansion, we can relate the trace term to connectivity motifs, in particular the cycles.  First, we have
\beq
\tr((I-h(s)W)^{-1})=\sum_{n=0}^{\infty} h(s)^n \tr(W^n) 
\label{E:trace_G}
\eeq
Note that $N^{-n}\tr(W^n)$ is the frequency with which an $n$-cycle of connections occurs in a network in which the entries of $W$ take values in $\{0,1\}$. A 2-cycle is simply a pair of reciprocal connections.
In general, a $n$-cycle is a loop identified with indices connected as $i_1\rightarrow i_2 \rightarrow \cdots \rightarrow i_n \rightarrow i_1$. Similar to the chain motifs considered earlier, we may define the motif moments for $n$-cycles as 
\beq
\mu^c_n=N^{-n}\tr(W^n) \; \; \; , \; \; \; n \ge 1.
\label{E:mu_cycle}
\eeq 
By generalizing the decomposition between motif moments and cumulants (Eq.~\eqref{E:mu_kappa}), we should expect that 
\beq
\mu^c_n=\sum_{\{n_1,  \cdots , n_t\} \in \mathcal{C}(n)} 
\frac{n}{t} \left(\prod_{i=1}^{t} \kappa_{n_i}\right) + \kappa^c_n.
\label{E:mu_kappa_cycle}
\eeq
The formula is explained by enumerating all possible decompositions of a $n$-cycle. We do this in two steps.  First, we break the cycle at any one of the $n$ nodes, yielding a single chain.  Next, we break the resulting chain into smaller chain motifs, which correspond to ordered partitions as before. Since each decomposition which ends up with $t\ge 1$ components can be acquired by first breaking the cycle at any of the $t$ locations, it is redundantly counted $t$ times in the procedure above. This, together with the $n$ locations of the first break, explains the $n/t$ factor in Eq.~\eqref{E:mu_kappa_cycle}. Finally, the only exception to the procedure is the cycle itself without any breaks, which is the last term $\kappa^c_n$. Naturally, we recursively define $\kappa^c_n$ as the cumulant for cycle motifs using Eq.~\eqref{E:mu_kappa_cycle}. One can also prove that this definition has an equivalent matrix expression $\kappa^c_n=N^{-n}\tr((\Theta W)^n)$, $\Theta=I-e e^\tsp$.

We can now use the combinatorial properties of $\kappa_n$ and $\kappa^c_n$ to obtain a resummed formula for this trace term arise in correlated input output weights.

\begin{restatable}{thmrestate}{thmcycle}
\label{thm:trace_motif_resum}
\begin{align}
\nonumber
&\rho \sigma^2 \tr((I-h W)^{-1})h =\\
&\rho \sigma^2 \left( Nh+\sum_{n=1}^{\infty}N^nh^{n+1}\kappa^c_{n}
+ \frac{\sum_{n=1}^{\infty} n N^{n}h^{n+1} \kappa_{n}}{1-\sum_{n=1}^{\infty} N^{n}h^n \kappa_{n}}
\right),
\label{E:trace_resum}
\end{align}
provided the connection strength is sufficiently small so that the series above converges (the condition for this being $\abs{h(s)} \rho(\Theta W\Theta)<1$).
\end{restatable}

\begin{proof}
We start by expanding $(I-hW)^{-1}$ into a matrix power series and then substitute $\tr(W^n)$ in for $\mu_n^{c}$, using the definition Eq.~\eqref{E:mu_cycle}. Next, we use decomposition Eq.~\eqref{E:mu_kappa_cycle} to split the terms of $\mu_{*}^c$ into $\kappa_{*}^c$ (${}_* $ stands for an arbitrary length index).  This gives
\beqrn
&&\tr((I-h W)^{-1})=\tr(I)+\sum_{n=1}^{\infty}h^n\tr(W^n)\\
&=&N+\sum_{n=1}^{\infty}N^{n}h^n\mu^c_n
=N+\sum_{n=1}^{\infty}N^nh^n\kappa^c_{n}\\
&&+\sum_{n=1}^{\infty}N^nh^n \sum_{\{n_1,  \cdots , n_t\} \in \mathcal{C}(n)} 
\frac{n}{t} \left(\prod_{i=1}^{t} \kappa_{n_i}\right)  \; .
\eeqrn
The essential step is to resum the last term in the above expression. We introduce a (formal) series in complex variable $z$,
\[
f(z)=\sum_{n=1}^{\infty} z^n \sum_{\{n_1,  \cdots , n_t\} \in \mathcal{C}(n)} 
\frac{n}{t} \left(\prod_{i=1}^{t} \kappa_{n_i}\right).
\]
The original series can be obtained by setting $z=N h$ once the series is summed to a closed expression. Formally, or for $z$ in the radius of convergence of the series, consider the indefinite integral of $f(z)/z$,
\beqrn
&&\int \frac{f(z)}{z} dz = \sum_{n=1}^{\infty} z^n \sum_{\{n_1,  \cdots , n_t\} \in \mathcal{C}(n)} 
\frac{1}{t} \left(\prod_{i=1}^{t} \kappa_{n_i}\right)\\
&=& \sum_{n=1}^{\infty} \sum_{\{n_1,  \cdots , n_t\} \in \mathcal{C}(n)} 
\frac{1}{t} \left(\prod_{i=1}^{t} z^{n_i} \kappa_{n_i}\right)\\
&=& \sum_{t=1}^{\infty} 
\frac{1}{t} \prod_{i=1}^{t} \left( \sum_{n_i=1}^{\infty} z^{n_i} \kappa_{n_i}\right)\\
&=&\sum_{t=1}^{\infty} 
\frac{1}{t} \left( \sum_{n=1}^{\infty} z^{n} \kappa_{n}\right)^t =-\log\left(1-\sum_{n=1}^{\infty} z^{n} \kappa_{n}\right).\\
\eeqrn
For the third ``=", we have used the same trick of switching the order of summations as in the proof of Theorem~\ref{TH:uniform_motif}, by enumerating $t$ first.
In the last equality, we treat $\sum_{n=1}^{\infty} z^{n} \kappa_{n}$ as a variable and assume its magnitude is less than 1 (or operate formally). Finally, 
\[
f(z)=-z \frac{\partial }{\partial z} \log\left(1-\sum_{n=1}^{\infty} z^{n} \kappa_{n}\right)
=\frac{\sum_{n=1}^{\infty} n z^{n} \kappa_{n}}{1-\sum_{n=1}^{\infty} z^{n} \kappa_{n}}.
\]
The original trace term for $G(s)$ is thus
\beqrn
&&\rho \sigma^2 \tr((I-h W)^{-1})h\\
&&=\rho \sigma^2 \left( Nh+\sum_{n=1}^{\infty}N^nh^{n+1}\kappa^c_{n} + \frac{\sum_{n=1}^{\infty} n N^{n}h^{n+1} \kappa_{n}}{1-\sum_{n=1}^{\infty} N^{n}h^n \kappa_{n}}
\right).
\eeqrn

\end{proof}

This expression Eq.~\eqref{E:trace_resum} again corresponds to a feedback diagram as shown in Fig.~\ref{fig:cycle_diagram}. Compare with the diagram (Fig.~\ref{fig:ladder_diagram}) in the uniform input output case,  note that the additional feedbacks by the cycle motif cumulants $\kappa_n^c$, as well as the different coefficient for the chain motif cumulant links. 

\begin{figure}[h]
\begin{center}
\includegraphics[width=0.23\textwidth]{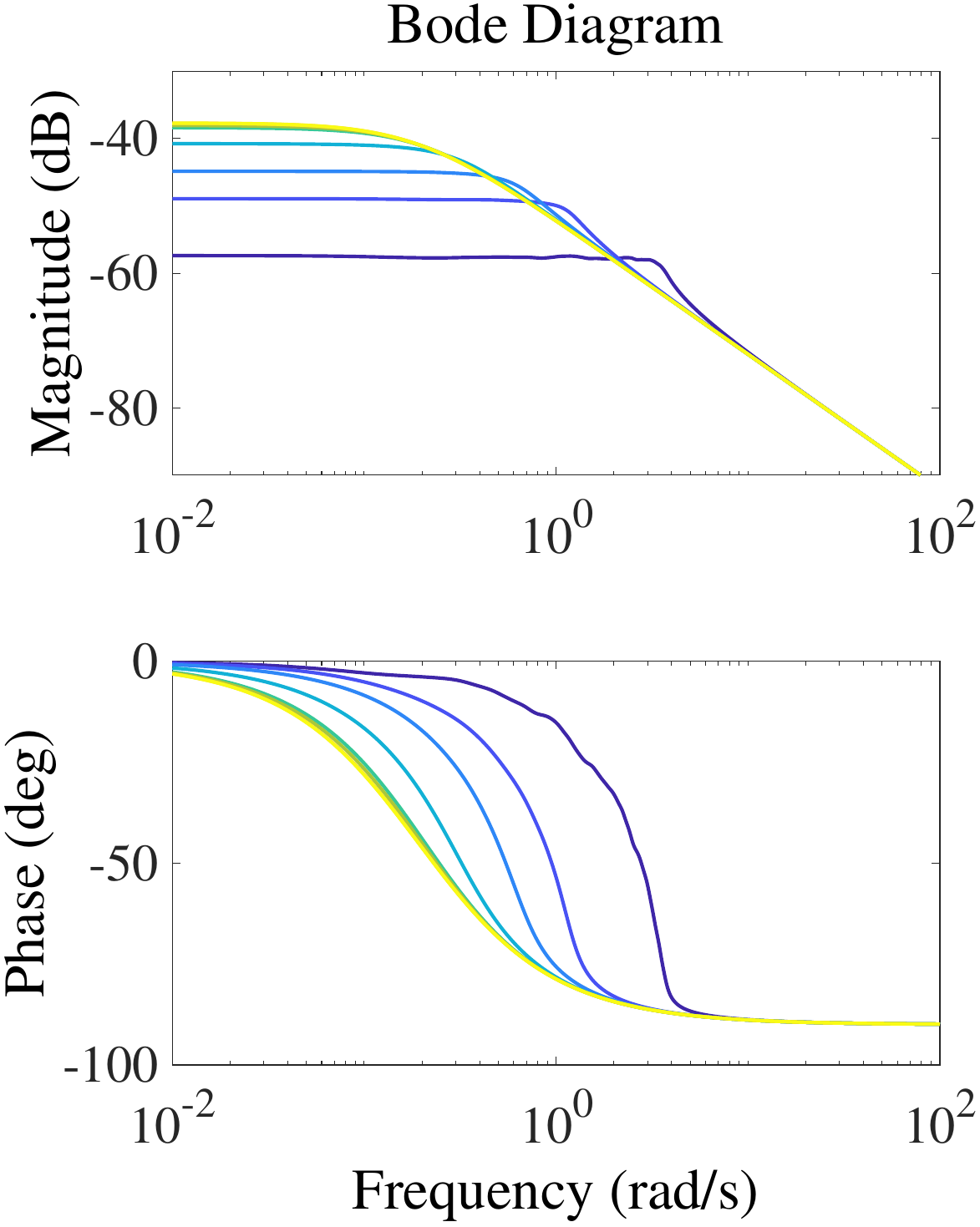}
\includegraphics[width=0.23\textwidth]{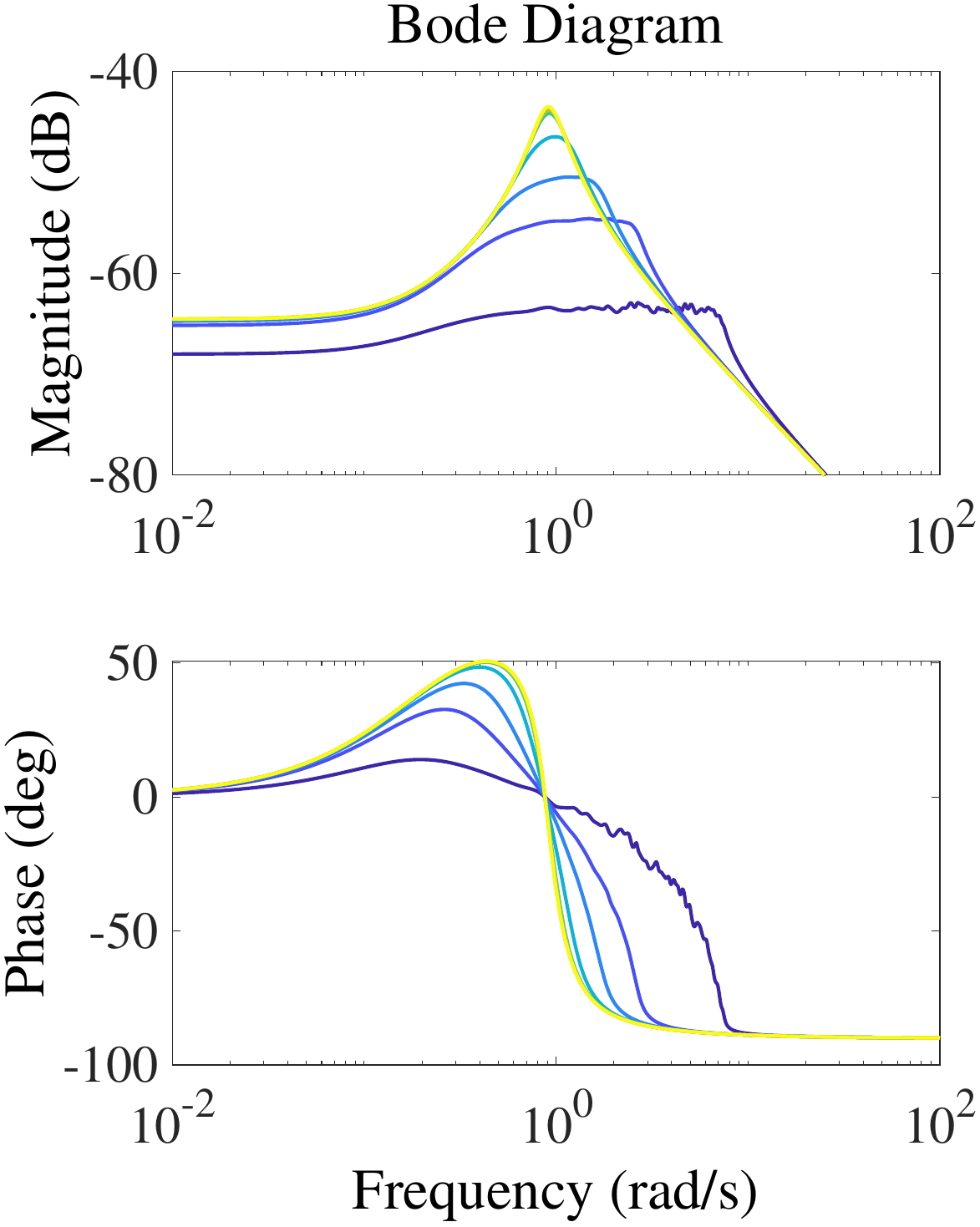}
\caption{Numerical examples showing the impact of $\kappa^c_2$ on $\EVs{G(s)}$ for correlated input and output weights $B,C$, for the node filter $h_{\exp}(s)$ (left column) and decaying-oscillatory filter $h_{\cos}(s)$ (right column) calculated using Eq.~\eqref{E:E_Gs_corr}. Lines of blue-to-yellow colors (darker to lighter shades) correspond to $\kappa_2^c/0.09$ taking values from -0.85 to 0.9. The entries of $B$ and $C$ have mean 0 and $\rho\sigma^2=1/N$ (see text). The networks $W$ are generated as Gaussian random matrices with entries of variance $0.09$ and size $N=400$ (Appendix~\ref{S:gaussian_cycle}). We scale the coupling strength of each $W$ to 40\% of the critical maximum value.}
\label{F:EG_corr}
\end{center}
\end{figure}

We illustrate the potential impacts of cycle motif cumulants on  $G(s)$ with the following numerical example. We use Eq.~\eqref{E:E_Gs_corr} to calculate  $\EVb{G(s)}$ while we have set $\theta=0$ and $\rho \sigma^2=1/N$. 
We generate networks based on Gaussian random entries with various levels of the cycle cumulant $\kappa^c_2$ (ranging from -0.85 $\sigma_w^2$ to 0.9 $\sigma_w^2$ whereas the maximum possible range is $[-\sigma_w^2,\sigma_w^2]$), and their $\EVb{G(s)}$ are plotted in Fig.~\ref{F:EG_corr} in lines of different colors.  The most significant impact of cycle motifs on the network transfer function happens with negative $\kappa^c_2$.  Overall, for the exponential node filter, positive $\kappa^c_2$ tends to increase the time constant; the opposite is true for negative $\kappa^c_2$.

Beyond calculating the network transfer function using the full connectivity $W$  via Eq.~\eqref{E:E_Gs_corr}, the resumming formula~\eqref{E:trace_resum} can be used to calculate $\EVs{G(s)}$ based only on the value of $\kappa^c_2$ (other motif cumulants are small in the generated networks).  We find that this approach is accurate for relatively small coupling strengths, and consequently the effect of $\kappa^c_2$ appears to be similar but smaller in scale than that shown in Fig.~\ref{F:EG_corr}.  For large coupling strengths,  however, the spectral radius condition required for using the motif cumulant expression (Eq.~\eqref{E:trace_resum}) is no longer satisfied. Extending our theory to capture such cases of strong coupling is left for future work.

\section{Networks with multiple populations and patterned inputs and outputs}
\label{S:multi_pop}

Above, while we allowed for randomness in the weights $B_i$ and $C_i$ by which a signal is read into or out of a network, we still assumed a single distribution for these weights; moreover, we took the single-node dynamics and  motif statistics to be homogeneous across the network.  Many networks of interest in biology and other fields, however, are composed of nodes of different types and connectivity rules (as for the proliferation of cell types in neuroscience). Here we describe how to generalize our theory to allow for dynamics ($h(s)$) and connectivity ($\kappa_n$) factors and input output weights ($B,C$) to be node-type specific, so the whole network is composed of multiple populations.

\begin{figure}[h]
\begin{center}
\includegraphics[width=0.35\textwidth]{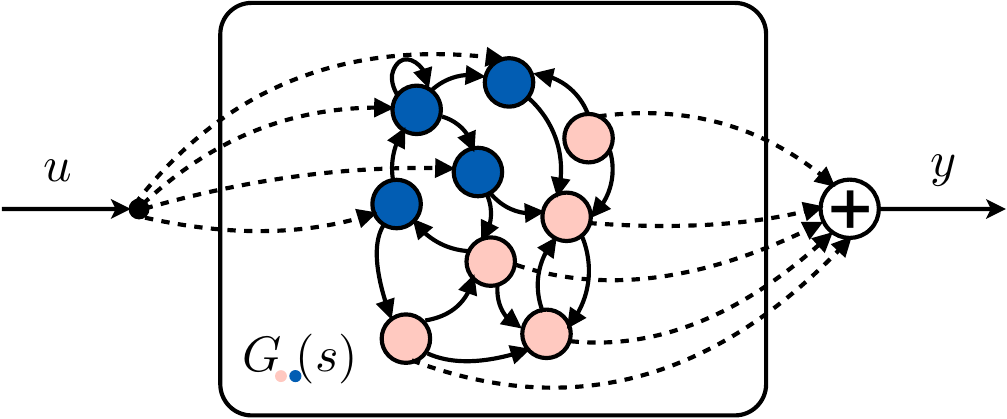}
\caption{Schematic for a network with multiple node types or population-wise patterned inputs/readouts; here, indicated by blue and pink node (dark and light shades).}
\label{fig:multi_pop_diagram}
\end{center}
\end{figure}

Consider a network consisting of $k$ populations of nodes, with population type indexed by $\alpha$. Nodes in each population $\alpha$ are assumed to have the same population specific filter $h_\alpha(s)$.  Ordering individual nodes in blocks according to their population index, the input-output equation for $x(s)$ (Eq.~\eqref{E:x_laplace_matrix}) can be expressed in matrix form by introducing the diagonal matrix $D_h=\diag(h_1,\cdots,h_1,h_2,\cdots,h_2,\cdots,h_k,\cdots,h_k)$:
\beq
x(s)=D_h(Wx(s)+Bu(s)).
\label{E:x_laplace_matrix_multi}
\eeq
Given the differences of node types, we consider signal input and output weights  $B,C$ that are different across populations yet uniform within nodes of a same population (Fig.~\ref{fig:multi_pop_diagram}). 

Any resulting network transfer function from these population specific weights is a linear combination of terms we denote by $G_{\alpha \beta}(s)$.  These are the network transfer functions achieved by uniformly feeding input the $u(s)$ into nodes of the population $\beta$ and reading the response out from nodes in population $\alpha$.
These transfer functions $G_{\alpha\beta}(s)$ form a matrix; by abuse of notation we refer to this matrix again by $G(s)$.  To derive a formula for it, first let $U$ be the block matrix 
\beqrn
\nonumber
&&U=\frac{1}{\sqrt{N}}
\left[ 
\barr{cccc} 
e_1 & 0 & \cdots & 0\\
0 & e_2 & \ddots & \vdots\\
\vdots & \ddots & \ddots & 0\\
0 & \cdots & 0 & e_k
\earr
\right], \\
&& e_\alpha=(1,\cdots,1)^\tsp \text{ (length $N_\alpha$)}.
\eeqrn
Here $N_\alpha$ is the size of population $\alpha$. We can then write $G(s)$ as
\beq
G(s)=U^\tsp (I-D_h W)^{-1} D_h U.
\label{E:transfer_G_multi}
\eeq
Note that $D_h$ and $U$ ``commute" due to their matching block structure, that is 
\[
D_h U=U D_h^\prime,\quad \text{where } D_h^\prime=\diag(h_1,\cdots,h_k).
\]
For simplicity, we will use the notation $D_h$ to represent $D_h^\prime$ whenever the meaning is clear from the dimensions of matrices.

Similarly as for Theorem~\ref{TH:uniform_motif}, we can rewrite the multipopulation network transfer function $G(s)$ in terms of \emph{population specific} motif cumulants that reflect node type identities in addition to connection structure. 
\begin{theorem}
\label{TH:uniform_motif_multi}
For a multipopulation network with dynamics satisfying Eq.~\eqref{E:transfer_G_multi}, the network transfer function can be written as
\beq
G(s)=\left(I-\sum_{n=1}^{\infty} N^n D_f\tilde{\kappa}_n \right)^{-1} D_f D_h
\label{E:transfer_motifs_multi}
\eeq
where $\tilde{\kappa}_n$ is the motif cumulant ($k\times k$ matrix) of $D_h W$ for length $n$ chains, defined via the recursive relations with (matrix) motif moments~\citep{Hu:2014gs},
\beq
\tilde{\mu}_n=\frac{1}{N^n} D_f^{-1}U^\tsp (D_h W)^n U D_f^{-1},
\label{E:define_mu_multi}
\eeq
\beq
\tilde{\mu}_{n} =
\sum_{\{n_1,  \cdots , n_t\} \in \mathcal{C}(n)} 
\left[ \left(\prod_{i=1}^{t-1} \tilde{\kappa}_{n_i} D_f \right) \tilde{\kappa}_{n_t} \right].
\label{E:mu_kappa_multi}
\eeq
Here $\mathcal{C}(n)$ is the set of all compositions (\emph{ordered} partitions) of $n$, and the diagonal matrix $D_f=\diag(N_1/N,\cdots,N_k/N)$.
\end{theorem}
Here we have combined the $h_\alpha$ filters with $W$ to define the motif statistics for the effective coupling matrix $\tilde{W}=D_h W$. This is indicated via the $\tilde{}$ over $\mu$ and $\kappa$. 

In the theorem above, we needed to introduce population specific motif moments $(\tilde{\mu}_n)_{\alpha \beta}$ and cumulants $(\tilde{\kappa}_n)_{\alpha \beta}$ (the entries of $\tilde{\mu}_n$ and $\tilde{\kappa}_n$). The meaning of $(\tilde{\mu}_n)_{\alpha \beta}$ is the frequency or probability of $n$-length chains with start and end nodes in populations $\beta$ and $\alpha$ respectively. The decomposition relation of population motif cumulants Eq.~\eqref{E:mu_kappa_multi} is a matrix version of Eq.~\eqref{E:mu_kappa},  and is formally identical if the multiplication of two matrix objects $A, B$ is implemented as $A D_f B$~\citep{Hu:2014gs}. The insertion of $D_f$ here provides the proper weights for averaging between populations with different sizes.

Moreover, the resummed motif cumulant expression of $G(s)$ Eq.~\eqref{E:mu_kappa_multi} looks almost identical to the single population case Eq.~\eqref{E:transfer_motifs}, and indeed can be proved in the same way, if we replace scalar quantities with $k\times k$ matrices (motif moments or cumulants) and again insert $D_f=\diag(N_1/N,\cdots,N_k/N)$ for matrix multiplications as described above.

\begin{figure}[h]
\begin{center}
\includegraphics[width=0.35\textwidth]{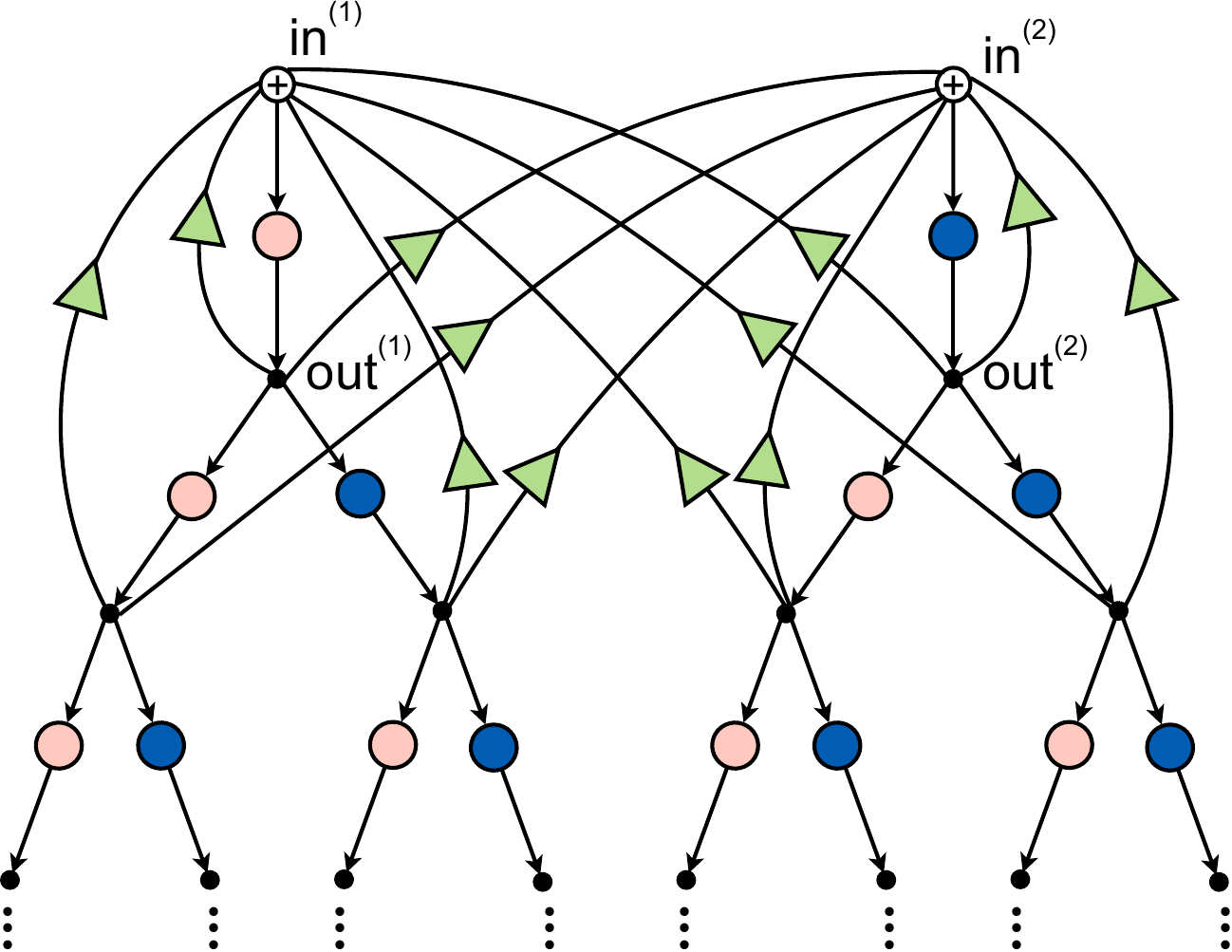}
\caption{Feedback diagram that gives the network transfer function matrix $G(s)$ for networks with two populations (different colors/shades). Various green triangles are feedbacks corresponding to population specific motif cumulants.}
\label{fig:tree_diagram}
\end{center}
\end{figure}

We can also directly express $\tilde{\kappa}_n$ in terms of motif cumulants of the original connectivity matrix $W$ as opposed to the filter-weighted matrix $D_h W$.  This will also lead to a multipopulation version of the feedback diagram (Fig.~\ref{fig:tree_diagram}). We illustrate this  for the case of two populations. Let $\kappa_1^{\alpha \beta}$ be the motif cumulant of $W$ for length 1 chains starting in population $\beta$ and ending in population $\alpha$. Similarly, let $\kappa_2^{\alpha \beta \gamma}$ be the motif cumulants of $W$ for length 2 chains with the three nodes in population $\gamma,\beta,\alpha$ respectively, and so on for higher order motifs.

By enumerating the population identity of nodes in chain motifs, it is easy to show that
\beq
D_f \tilde{\kappa}_1=
D_f D_h
\left[
\barr{cc}
\kappa_1^{11} & \kappa_1^{12}\\
\kappa_1^{21} & \kappa_1^{22}
\earr
\right],
\label{E:kappa_multi_1}
\eeq
\beqr
\nonumber
&&D_f\tilde{\kappa}_2=D_f D_h\\
&&\cdot
\left[
\barr{cc}
\frac{N_1}{N} h_1 \kappa_2^{111}+\frac{N_2}{N} h_2 \kappa_2^{121} 
& \frac{N_1}{N} h_1 \kappa_2^{112}+\frac{N_2}{N} h_2 \kappa_2^{122}\\
\frac{N_1}{N} h_1 \kappa_2^{211}+\frac{N_2}{N} h_2 \kappa_2^{221} 
& \frac{N_1}{N} h_1 \kappa_2^{212}+\frac{N_2}{N} h_2 \kappa_2^{222}
\earr
\right].
\label{E:kappa_multi_2}
\eeqr
The above formulae motivate defining $\tilde{h}_\alpha=\frac{N_\alpha}{N} h_\alpha$, $\alpha=1,2$, and $D_{\tilde{h}}=D_f D_h=\diag(\tilde{h}_1,\tilde{h}_2)$ to simplify the expressions. Using this notation, we can, for example, rewrite Eq.~\eqref{E:kappa_multi_1} and \eqref{E:kappa_multi_2}
as
\beq
D_f \tilde{\kappa}_1=
D_{\tilde{h}}
\left[
\barr{cc}
\kappa_1^{11} & \kappa_1^{12}\\
\kappa_1^{21} & \kappa_1^{22}
\earr
\right],
\label{E:kappa_multi_1_tilde}
\eeq
\beq
D_f\tilde{\kappa}_2=
D_{\tilde{h}}
\left[
\barr{cc}
\tilde{h}_1 \kappa_2^{111}+\tilde{h}_2 \kappa_2^{121} 
& \tilde{h}_1 \kappa_2^{112}+\tilde{h}_2 \kappa_2^{122}\\
\tilde{h}_1 \kappa_2^{211}+\tilde{h}_2 \kappa_2^{221} 
& \tilde{h}_1 \kappa_2^{212}+\tilde{h}_2 \kappa_2^{222}
\earr
\right].
\label{E:kappa_multi_2_tilde}
\eeq
Plugging Eqns.~(\ref{E:kappa_multi_1_tilde}-\ref{E:kappa_multi_2_tilde}) and analogous expressions at higher orders into Eq.~\eqref{E:transfer_motifs_multi} gives a series expression of $G(s)$ in terms of population specific motif cumulants such as $\kappa_2^{\alpha \beta \gamma}$.

Based on these calculations Eqns.~(\ref{E:kappa_multi_1_tilde}, \ref{E:kappa_multi_2_tilde}), we construct a feedback diagram for the formula in Theorem~\ref{TH:uniform_motif_multi} (Fig.~\ref{fig:tree_diagram}).
The feedback diagram consists of two (infinite) perfect binary trees, whose roots have an in and out node and a link between the two, carrying filters $\tilde{h}_\alpha=\frac{N_\alpha}{N} h_\alpha$. The rest of the trees grow from the out nodes. Each left branch has a filter $\tilde{h}_{1}$ and each right branch has a filter $\tilde{h}_2$. At every node of the binary trees (i.e. all nodes except for the two in nodes), there are two links connecting to nodes $\text{in}^{(1)}$ and $\text{in}^{(2)}$ respectively. The strength of such feedback links are determined by population specific chain motif cumulants as $N^{n}\kappa_n^{\alpha \text{path} \beta}$. Here ``path" in the superscript is the sequence with $1$ or $2$ denoting left or right branches traveled along the path from the root to this node (starting from the end of the sequence); $\beta$ is the index of the tree that the node belongs to, and $\alpha$ is the index of the in node that the link connects to. Sending input into one of the two in nodes and reading it out from one of the two out nodes gives the corresponding entry in the $2 \times 2$ matrix $G(s)$.

Finally, we can generalize the relation between the cut-off time constant and motif cumulants for networks with multiple node populations.
\begin{theorem}
\label{TH:time_constant_multi}
Assume that the node filter for a population $h_\alpha(s)$ decreases asymptotically as $1/s^{g_\alpha}$ for large $s$ and $g_\alpha > 0$, with a time constant $\tau_\alpha$. We form a ``scalar" transfer function based on the matrix of the network transfer functions $G(s)$ via a linear combination using vectors $\hat{B}$ and $\hat{C}$:
\[
\hat{G}(s)=\hat{C}^{\tsp} G(s) \hat{B}.
\]
Then the time constant of $\hat{G}(s)$ is
$\left(
\frac{\hat{C}^{\tsp}f(D_\tau)\hat{B}}  {\hat{C}^\tsp D_f \hat{B}}
\right)^{\frac{1}{g_0}}
$, where $f(D_\tau)$ is a matrix acquired by replacing the diagonal matrix $D_{h}$ in Eq.~\ref{E:transfer_motifs_multi} (and inside the definition of $\tilde{\kappa}_n$) by another diagonal matrix $D_{\tau}$,
\[
D_{\tau}=\diag\{\tau_{1}^{g_1},\ldots,\tau_{k}^{g_k}\},
\]
\[
f(D_\tau)=\left(I-\sum_{n=1}^{\infty} N^n D_f\tilde{\kappa}_n \right)^{-1} D_f D_\tau.
\]
The degree of asymptotic decay of $\hat{G}(s)$, $g$, is determined by both the vectors $\hat{B},\hat{C}$ and the motif cumulant structure of $W$. In the special case when $g_\alpha \equiv g_0$ are all equal, and $\hat{C}^{\tsp} D_f \hat{B}\neq 0$, we have $g=g_0$.
\end{theorem}

\begin{proof}
As in the proof of the single population case, the time constant of $\hat{G}(s)$ is determined by $\hat{G}(0)$ and its leading term of asymptotic decay with $s$.

For $\hat{G}(0)$,
\beq
\hat{G}(0)=\hat{C}^\tsp G(0) \hat{B}
\label{E:G_zero_multi}
\eeq
and we can evaluate $G(0)$ using Eq.~\eqref{E:transfer_motifs_multi}, which is simply replacing $D_h$ with 
\[
\diag (h_1(0),\ldots,h_k(0))=\diag (\tau_1^{g_1},\ldots,\tau_k^{g_k})=D_\tau
\]
in Eq.~\eqref{E:transfer_motifs_multi}. In other words, $G(0)=f(D_\tau)$, using the definition of $f(\cdot)$ given in the statement of the theorem.

To determine the leading decay term in the special case when the $g_\alpha=g_0$ are all equal, we can expand Eq.~\eqref{E:transfer_motifs_multi} in terms of powers of $h_\alpha$,
\beq
\hat{G}(s)=\hat{C}^{\tsp} G(s) \hat{B}=\hat{C}^{\tsp}(I+N D_f \tilde{\kappa}_1+\cdots )D_f D_h \hat{B}.
\label{E:expand_G_mult}
\eeq
Using the assumption that $\hat{C}^\tsp D_f \hat{B}\neq 0$, the first term in Eq.~\eqref{E:expand_G_mult}
\[
\hat{C}^\tsp D_f  D_h \hat{B} \approx \frac{1}{s^{g_0}}\hat{C}^\tsp D_f \hat{B} \quad \text{as } s \rightarrow \infty.
\]
The order of $s$, as $s\rightarrow \infty$, for all other terms in Eq.~\eqref{E:expand_G_mult} is at most $-2g_0$. Therefore, $G(s)$ decays as $\hat{C}^\tsp D_f \hat{B} /s^{g_0}$. Combining with Eq.~\eqref{E:G_zero_multi}, we conclude that the time scale of $\hat{G}(s)$ is 
$\left(
\frac{\hat{C}^{\tsp}f(D_\tau)\hat{B}}  {\hat{C}^\tsp D_f \hat{B}}
\right)^{\frac{1}{g_0}}
$.
\end{proof}

\section{Degree-corrected motif cumulants $\kpdeg_n$ that improve the description of the network transfer function}
\label{S:degree_corrected_motif}

In this section,  we develop a modified version of motif cumulants that allows for describing networks with strong degree heterogeneity, which is a common feature in many real world networks. This method -- which we call degree-corrected motif cumulants $\kpdeg_n$-- therefore significantly broadens the applications of our theory.
The $\kpdeg_n$, while slightly more complex than the original $\kappa_n$, can produce highly accurate approximations for $G(s)$ when the corresponding series is truncated to use only $\kpdeg_1$ and $\kpdeg_2$ --- even when this approach fails for the original $\kappa_n$. Importantly, the $\kpdeg_n$ can be expressed in terms of the original $\kappa_n$ through algebraic combinations, and thus are also local network statistics and  require no additional information (such as the degree distribution) about the connectivity. As an application, we use this degree-corrected motif cumulant theory to describe networks with stronger effects from motif cumulant $\kappa_2$ (Fig.~\ref{F:change_kappa2_exp} and \ref{F:change_kappa2_cos}).

We start from the following general result, but without going into details as its careful exposition will be the subject of future work. It is possible to carry out the resumming for Eq.~\eqref{E:transfer_G_BC} for arbitrary weights $B,\, C$ analogously to Eq.~\eqref{E:transfer_motifs}:
\beq
\Gbc(s)=\Nbc \left(1-\sum_{n=1}^{\infty} N^n \kpbc_{n} h^n(s)\right)^{-1} h(s).
\label{E:transfer_motifs_BC}
\eeq
Here $\Nbc=C^\tsp B$, $\Thbc:=I-\frac{1}{\Nbc}B C^\tsp$. We use notation $\Gbc(s)$ instead of $G(s)$ to signify the use of weights $B,\, C$. Unless stated otherwise, In this section $G(s) = G^{ee}(s)$  means the uniform weight case where $B=C=l= (1,\ldots,1)^T$ (it differs from previously by a constant $N$). The key ingredient in Eq.~\eqref{E:transfer_motifs_BC} are weighted  ``motif moments" $\mubc_n$ and ``motif cumulants"  $\kpbc_n$, defined as
\beqrn
\nonumber
\mubc_n &=&\frac{1}{N^{n}\Nbc}C^\tsp W^n B,\\
 \kpbc_{n}&=&\frac{1}{N^{n}\Nbc}C^\tsp W(\Thbc W)^{n-1} B.
\eeqrn
Importantly, the same decomposition relation Eq.~\eqref{E:mu_kappa} also holds for $\mubc_n$ and $\kpbc_n$.

A special case of the generalized resumming formula is to choose $B,\, C$ as in- and out- degrees of $W$, that is, 
\[
B=W l, \quad C^\tsp =l^\tsp W.
\]
This choice of $B,\, C$ is in order to form a most efficient weighted motif cumulant series in the resummed formula Eq.~\eqref{E:transfer_motifs_BC}, and motivated by observations and heuristic arguments about eliminating the dominant eigenvalue of $W$ in $\Theta W\Theta$, based on the degree vector approximating the \PF{}  vector for $W$.  This is partially discussed in~\citep{Hu:2014gs}, and again we will save exposition of further details for future work. 

Another advantage of choosing $B,C$ as degrees is that Eq.~\eqref{E:transfer_motifs_BC} can be directly related to the original $G(s)$ and regular motif cumulants, thus requiring no additional information of the network connectivity. To signify this special choice of $B,\,C$, we will use the notation $\Gdeg(s)$, $\mudeg_n$, $\kpdeg_n$ and $\Ndeg$ as a special case for $\Gbc(s)$, $\mubc_n$, $\kpbc_n$ and $\Nbc$.

For general $W$ (with non-uniform degrees), the network transfer function $\Gdeg(s)$ given by Eq.~\eqref{E:transfer_motifs_BC} is different from $G(s)$, the latter being based on uniform input and output weights. Nonetheless, these are related:
\beqrn
\Gdeg(s)&=&C^\tsp (I-h(s)W)^{-1} B h(s)\\
&=& l^\tsp W(I-h(s)W)^{-1} W l h(s)\\
&=& l^\tsp \sum_{n=2}^{\infty} h(s)^{n-2}W^{n} l h(s)\\
&=& \frac{N}{h^2(s)} \left( G(s)-h(s)-N \kappa_1 h^2(s)\right) .
\eeqrn
Therefore
\beq
G(s)=\frac{1}{N} h^2(s) \Gdeg(s) +h(s)+N \kappa_1 h^2(s).
\label{E:G_GBC}
\eeq

The next step is to write $\Gdeg(s)$ in terms of the original $\kappa_n$. Given Eq.~\eqref{E:transfer_motifs_BC}, this boils down to expressing $\kpdeg_n$ in terms of $\kappa_n$. The basic relationship is through the motif moments $\mudeg_n$ and $\mu_n$,
\beq
\mudeg_n= \frac{1}{N^{n+1}}\frac{l^\tsp l} { C^\tsp  B}C^\tsp W^n B=\frac{ \mu_{n+2}}{\mu_2}.
\label{E:mubc_mu}
\eeq
Eq.~\eqref{E:mubc_mu} has an intuitive explanation: $\mudeg_n$ is probability of length $n$ chains when each count is weighted by the  out-degree of the ``sending" node times the in-degree of the ``receiving" node.

\begin{figure}[h]
\begin{center}
\includegraphics[width=0.25\textwidth]{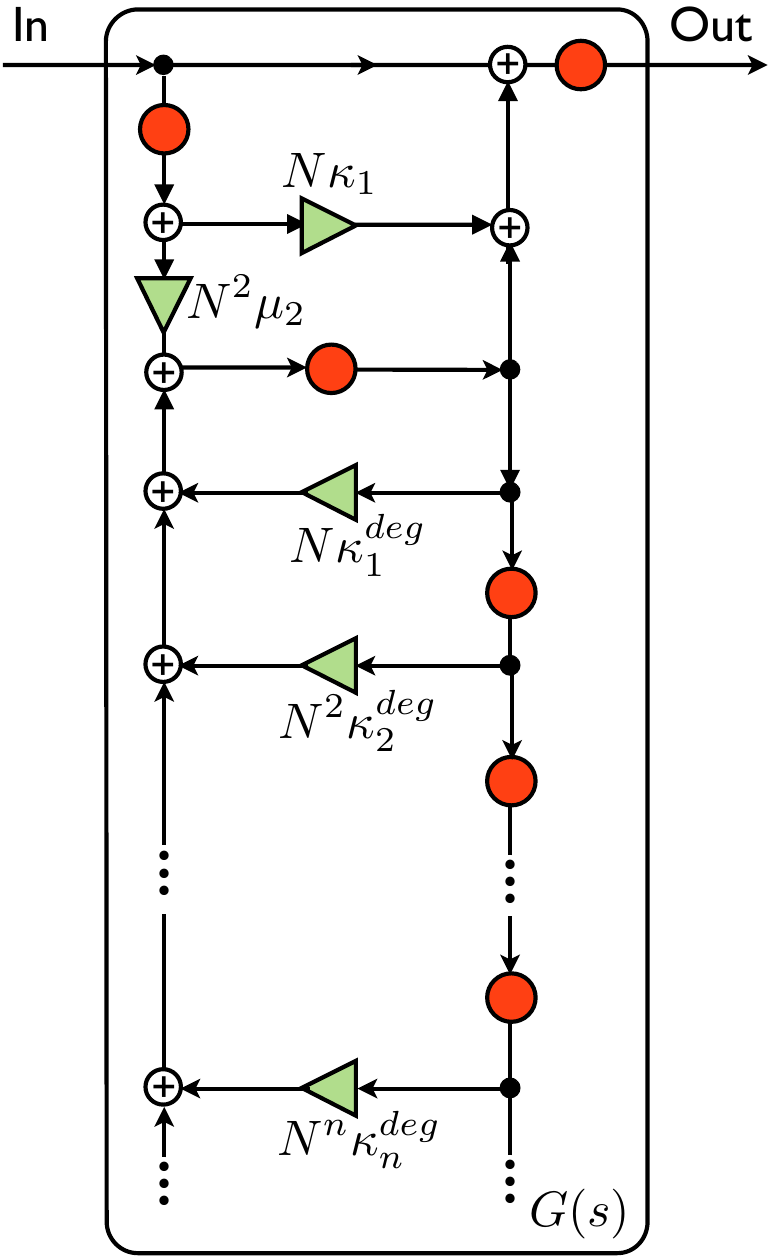}
\caption{Feedback diagram for the expression of $G(s)$ using degree-corrected motif cumulants $\kpdeg_n$.}
\label{fig:diagram_degree_corrected}
\end{center}
\end{figure}

Using Eq.~\eqref{E:mubc_mu} and the combinatorial relation Eq.~\eqref{E:mu_kappa} (same for degree-corrected motifs), we can derive the needed relationship among motif cumulants. For example,
\beqrn
\kpdeg_1 &=& \frac{\mu_3}{\mu_2}
=\frac{\kappa_3+2\kappa_2\kappa_1+\kappa_1^3}{\kappa_2+\kappa_1^2}\\
\kpdeg_2 &=& \mudeg_2-(\mudeg_1)^2
=\frac{\mu_4}{\mu_2}-\left(\frac{\mu_3}{\mu_2}\right)^2\\
&=&\frac{\kappa_4\kappa_2-2\kappa_3\kappa_2\kappa_1+\kappa_2^3
+\kappa_4\kappa_1^2-\kappa_3^2}{\mu_2^2}\\
\kpdeg_3 &=& \mudeg_3-2\mudeg_2\mudeg_1+(\mudeg_1)^3\\
&=& \frac{\mu_5}{\mu_2} -2\frac{\mu_4\mu_3}{\mu_2^2} +\left(\frac{\mu_3}{\mu_2}\right)^3\\
&=&\frac{1}{\mu_2^3}(\kappa_5\kappa_1^4+\kappa_5\kappa_2^2+2\kappa_5\kappa_2\kappa_1^2
+2\kappa_3^2\kappa_2\kappa_1\\
&&+3\kappa_3\kappa_2^2\kappa_1^2+\kappa_3^3
-2 \kappa_4 \kappa_2\kappa_1^3  -2\kappa_4\kappa_3\kappa_1^2\\
&&-2\kappa_4\kappa_2^2\kappa_1
-2\kappa_4\kappa_3\kappa_2
- \kappa_3^2\kappa_1^3-\kappa_2^4 \kappa_1)
\eeqrn
Using these expressions for $\kpdeg_n$ together with Eq.~\eqref{E:transfer_motifs_BC} and \eqref{E:G_GBC}, we can express $G(s)$ in terms of $\kappa_n$, and similarly draw a feedback diagram representation (Fig.~\ref{fig:diagram_degree_corrected}). The lower part of the diagram, beginning with the link $\kappa_1^{deg}$, corresponds to $\Gdeg(s)$ and has essentially the same structure as the original diagram Fig.~\ref{fig:ladder_diagram}.  The remaining few links on the top correspond to the additional terms in Eq.~\eqref{E:G_GBC}. Despite differences to Eq.~\eqref{E:transfer_motifs}, the two expressions are equivalent, in the sense that when being expanded as an infinite series in powers of $h(s)$, the coefficients containing $\kappa_n$ should be the same.  

The real difference (and advantage) of Eq.~\eqref{E:G_GBC} can be understood as a re-ordering of the terms in an infinite series. The re-ordered series may have an different convergence region (in terms of $h(s)$), and differ in value for finite truncations.  This latter property is what the leads to the improved approximations of the network transfer function based on the statistics of small motifs alone.

\subsection{Networks with stronger impact from $\kappa_2$}
\label{sec:stronger_k2_effect}
\begin{figure}[h]
\begin{center}
\includegraphics[width=0.45\textwidth]{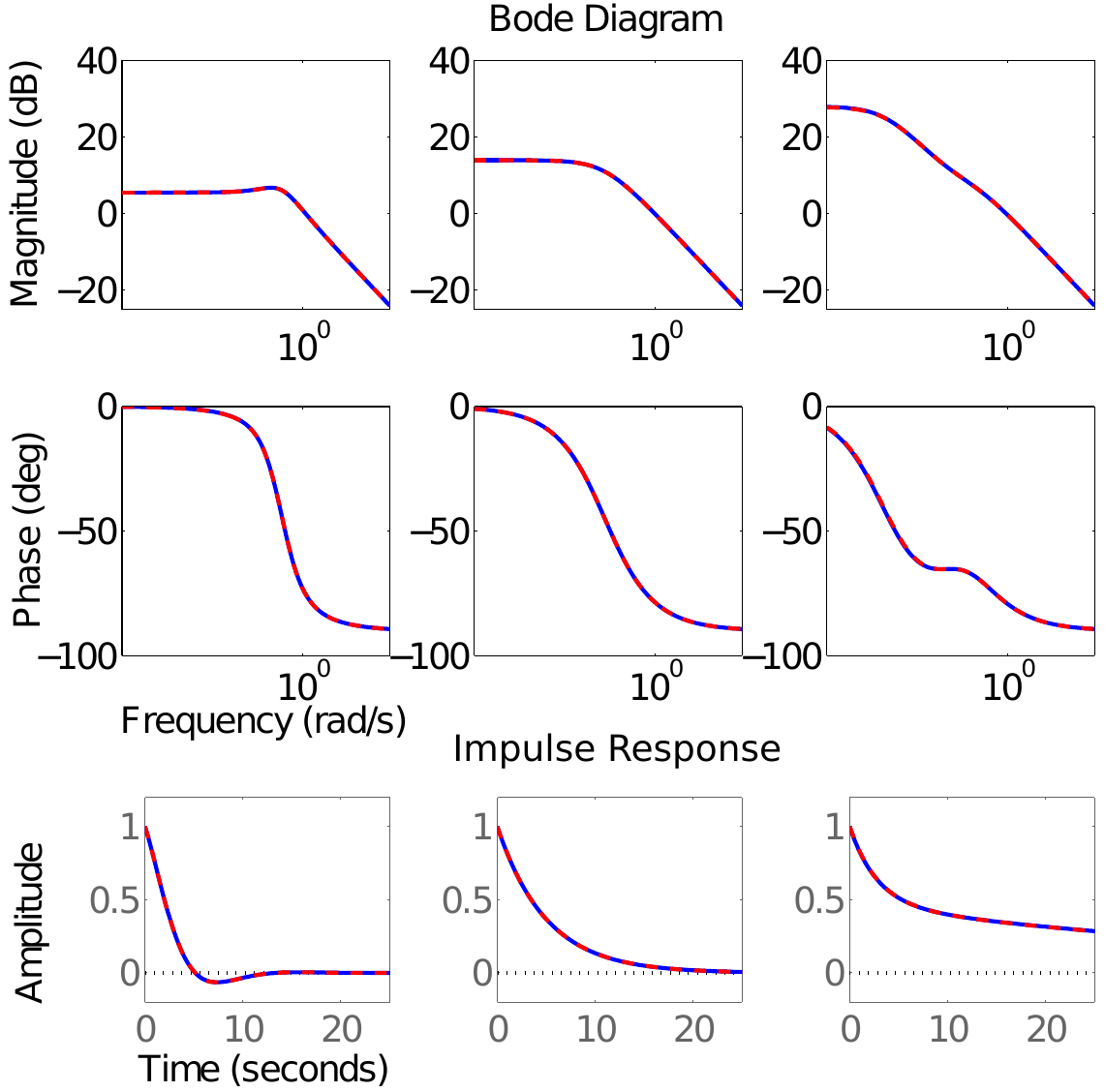}
\caption{Stronger effect of the length-2 chain motif cumulant $\kappa_2$ on shaping the network transfer function $G(s)$, for networks with node filter $h_{\exp}(s)$. The networks $W$'s are generated from the SONET random graph model (Appendix~\ref{S:sonet}).  Network size is $N=1000$, connection probability $\kappa_1=0.1$, and $\kappa_2/\kappa_1^2=-0.6,0,0.1$ respectively for the three columns from left to right. Higher order motif cumulants $\kappa_{n\ge 3}$ are small (see Appendix \ref{S:numerical_detail}). To emphasize the effect of $\kappa_2$, a global (negative) feedback is applied to all three networks to shift $\kappa_1$ to 0 without changing higher order motif cumulants. The blue solid lines are calculated by directly solving the system Eq.~\eqref{E:transfer_G_BC} using entire connectivity matrix $W$. The red dashed lines are calculated using only the first two degree-corrected motif cumulants $\kpdeg_1$ and $\kpdeg_2$ (see Sec.~\ref{S:degree_corrected_motif}), along with a formula analogous to Theorem \ref{TH:uniform_motif}.  The same connection strength constant multiplies $W$ for the three networks. The value of this constant is set to be 90\% of the maximum value under which all three networks are stable.
The parameters used are detailed in Appendix~\ref{sec:node_filter}. }
\label{F:change_kappa2_exp}
\end{center}
\end{figure}

\begin{figure}[h]
\begin{center}
\includegraphics[width=0.45\textwidth]{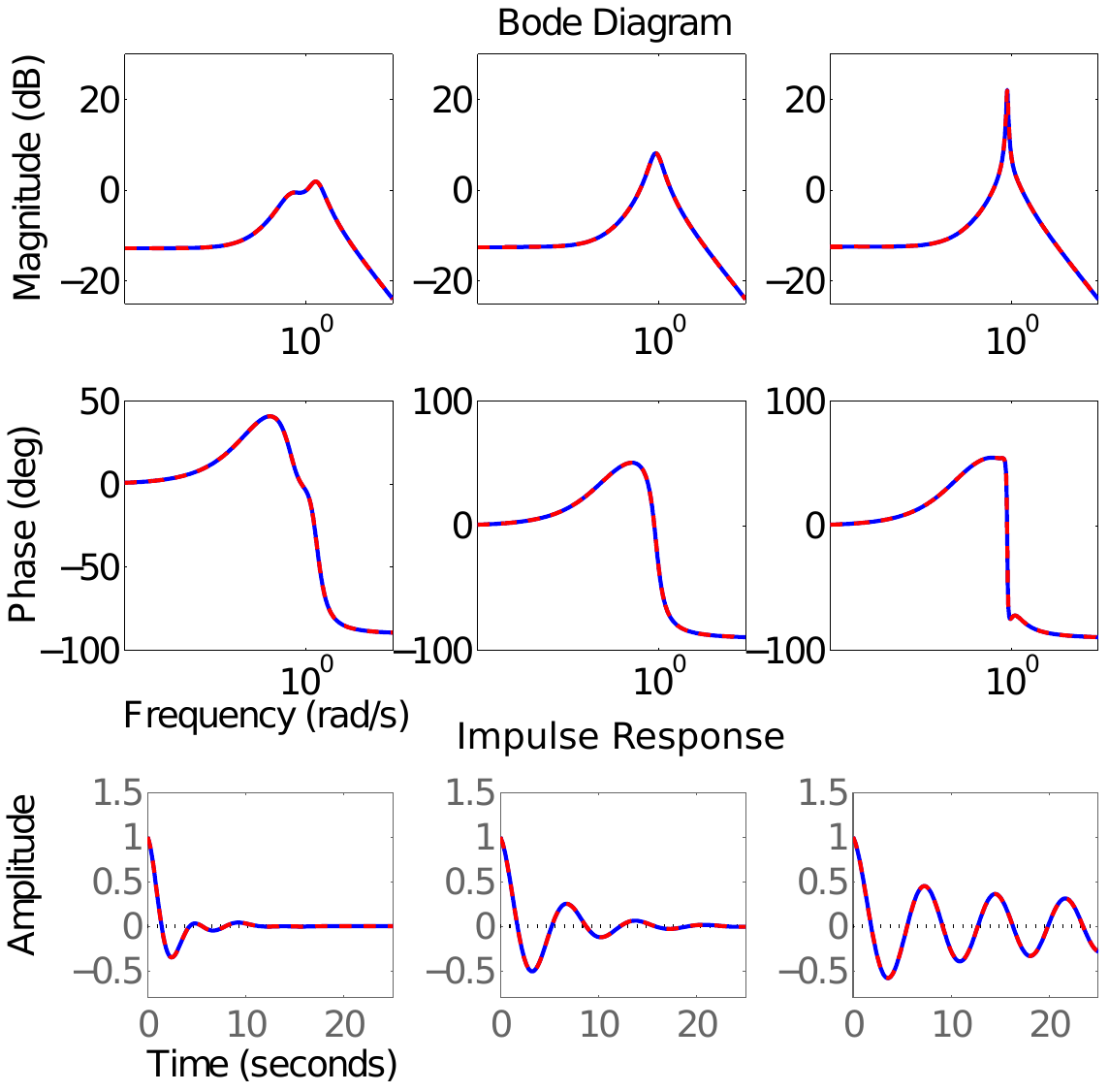}
\caption{Same as Fig.~\ref{F:change_kappa2_exp}, but for networks with node filter being $h_{\cos}(s)$.}
\label{F:change_kappa2_cos}
\end{center}
\end{figure}

Here we consider similar examples of complex networks as in Sec.~\ref{sec:change_kappa2}, but with stronger effects from these higher order motif cumulants.
To achieve this, we vary the motif cumulant $\kappa_2$ while leaving the other motif cumulants as close to zero as possible. One way to do this is to generate Gaussian networks as described in Appendix~\ref{S:network_generation}. In this case, $\kappa_1=0$ as the entries have zero mean; and higher order cumulants $\kappa_{n\ge 3}=0$ thanks to properties of Gaussian random variables. The numerical effect of $\kappa_2$ for these Gaussian networks is similar to the examples discussed below (Fig.~\ref{F:change_kappa2_exp} and~\ref{F:change_kappa2_cos}).

Note that Gaussian networks are \emph{densely connected}, in the sense that nodes are all to all connected with continuously distributed connection strengths. We can also  achieve similar effects for ``sparsely connected" networks where only a fraction of the connections are non-zero. We accomplish this in two steps.  First, to set $\kappa_{n\ge 3}$ small (but non-zero), we draw networks from the graph model of second order networks (SONETs~\citep{Zhao:2011dv}), which generalizes the \ER{} graph  by allowing non-zero $\kappa_2$ and produces a sparse, binary $W$ (it degenerates to a ER graph when $\kappa_2=0$).  Next, to set $\kappa_1=0$, we apply a global feedback of $-N\kappa_1 y(t)$.  As explained in Sec.~\ref{sec:proportional_feedback}, this will set the effective $\kappa_1=0$ while keeping all other $\kappa_n$ the same, and is equivalent to adding a constant of $-\kappa_1$ to all $W_{ij}$.  

The standard motif cumulant theory Eq.~\eqref{E:transfer_motifs} that works well for the Gaussian networks and the SONETs with small coupling strength starts to break down at strong coupling for SONETs. Higher order motif terms have to be included in Eq.~\eqref{E:transfer_motifs} to generate a good approximation of $G(s)$ for the positive $\kappa_2$ networks (need up to $\kappa_5$ terms to achieve the accuracy similar to the dashed lines in Fig.~\ref{F:change_kappa2_exp} and \ref{F:change_kappa2_cos}). Moreover, for the case of $\kappa_2<0$, keeping more terms in Eq.~\eqref{E:transfer_motifs} will not improve the approximation and can even make it worse! The problem is that the condition in Theorem~\ref{TH:uniform_motif} about the spectral radius is no longer satisfied, and the infinite series of motif cumulants in the denominator of Eq.~\eqref{E:transfer_motifs} diverge.

These difficulties for describing strong motif cumulant effects for ``sparse" networks can be resolved by using the degree-corrected motif cumulant theory that we developed above. By truncating Eq.~\eqref{E:transfer_motifs_BC} after $\kpdeg_2$, we achieve very accurate approximations to $G(s)$ (red dashed lines in Fig.~\ref{F:change_kappa2_exp} and~\ref{F:change_kappa2_cos}). 

Figures~\ref{F:change_kappa2_exp} and~\ref{F:change_kappa2_cos} show the impact of varying $\kappa_2$ for two different nodal filters  $h_{\exp}(s)$  (exponential decay) and $h_{\cos}(s)$ (oscillating decay) respectively. 
First, the middle columns show the case with all $\kappa_n$ very close to 0: the network used here is an \ER{} network and a global feedback is applied to shift $\kappa_1$ to 0.  We therefore recover the original node filter $G(s)=h(s)$.  Next, the right columns show that increasing $\kappa_2$ to positive values has a roughly similar effect on network transfer functions as increasing $\kappa_1$, for both of the filters. 

Intriguingly, however, the left columns show that networks with negative $\kappa_2$ --- fewer chain motifs --- produce qualitative changes in the network transfer functions (not seen in the networks with weaker $\kappa_2$ effect). For the exponential node filter $h_{\exp}(s)$ (Fig.~\ref{F:change_kappa2_exp}), a frequency peak is generated the Bode plot for response magnitudes. For the decaying-oscillatory filter $h_{\cos}(s)$ (Fig.~\ref{F:change_kappa2_cos}), the original resonant peak splits into two peaks. As $\kappa_2$ becomes more negative still, the peak seen for $h_{\exp}(s)$ increases its magnitude, and the twin peaks for $h_{\cos}(s)$ become more separated (data not shown). The effects of chain motif cumulants are also reflected in the impulse response functions.  In particular, note the emergence of a negative response window for $h_{\exp}(s)$ node filters, and irregular-looking oscillations in the impulse response for $h_{\cos}(s)$.

\section{Applications to real world networks}
\label{sec:real_networks}
\subsection{Mouse brain connectivity}
\begin{figure}[h!]
\begin{center}
\includegraphics[width=.48\textwidth]{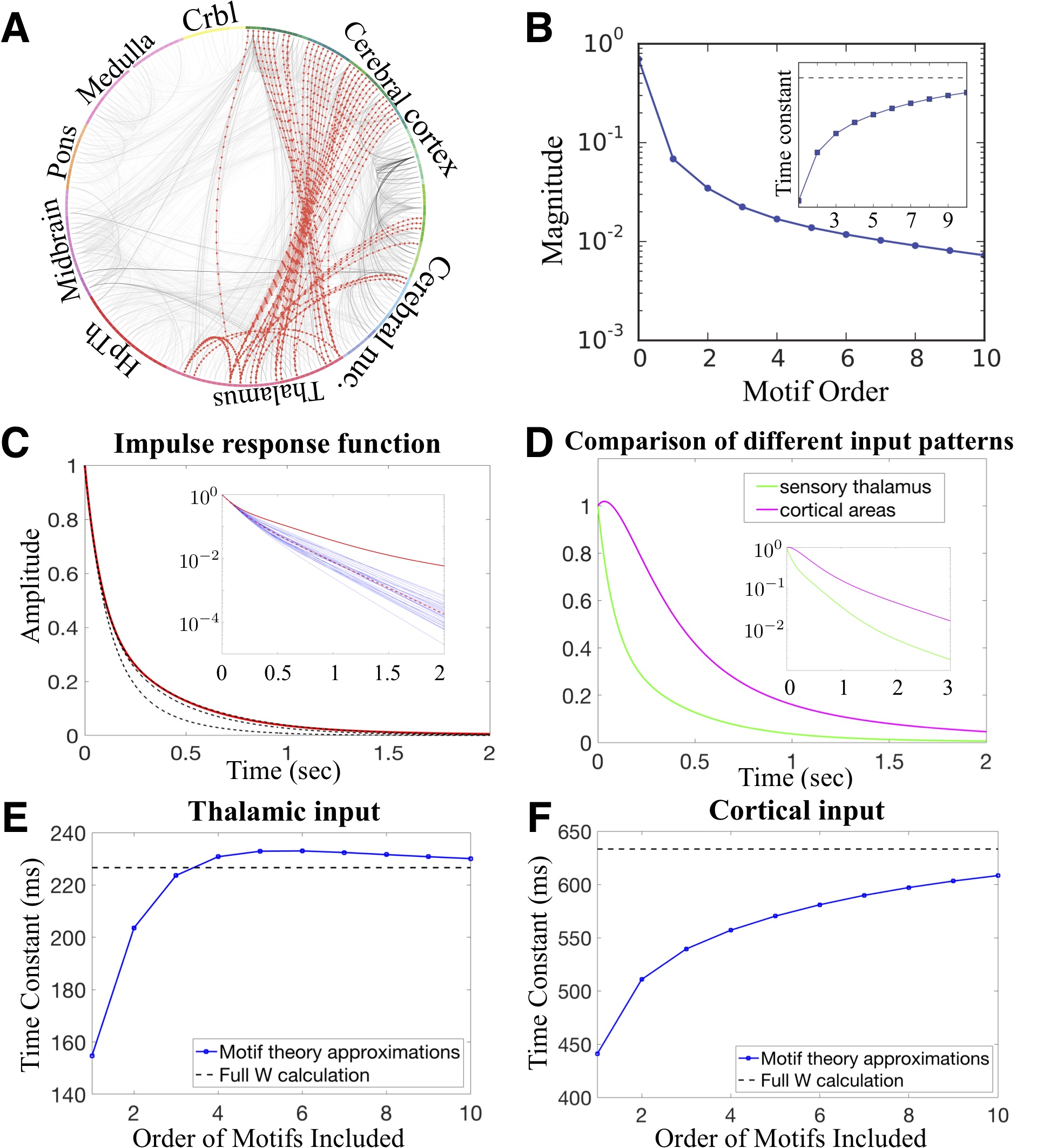}
\caption{{\bf Mesoscale mouse brain network dynamics.} {\bf A}. Mouse brain mesoscale  connectivity organized according to anatomical groups (Crbl: cerebellum, HyTh: hypothalamus, nuc.: nuclei); red dotted connections originate from 11 sensory thalamic nuclei; remaining connections are shaded proportional to connection strength, from~\cite{oha2014}. 
{\bf B}. The network-wide motif cumulants magnitude (Appendix \ref{sec:motif_magnitude_scaling}) of the original network do not decay rapidly with order (size), resulting in a long timescale of the network transfer function (inset) for a global input and readout. 
{\bf C}. The impulse response function (red solid line, main and inset) exhibits multiple timescales. The input signal is sent uniformly to sensory thalamic nuclei and read out from all regions. Black dashed lines depict successive (improving) approximations to this response computed by considering additional motif cumulants up to order 3. Inset: When the network is modified by either performing a node-degree preserving shuffle (thin-blue lines), or by setting every connection to a strength equal to the mean of the original log-normal weight distribution (red-dashed line), the long-time memory capacity of the network is diminished.
{\bf D}. 
Comparing timescales of the whole-brain responses to cortical signaling from higher level cortical areas (magenta, upper curve) and to input from sensory thalamus (green, lower curve, same as in C). Inset is plotting in log-y scale.
{\bf E}. 
Estimations of time constant for the thalamic input response (panel C) by successively including higher order motif cumulants.  
{\bf F}. 
Same as E except input is sent through higher level cortical areas.
}
\label{fig:mouse_brain}
\end{center}
\end{figure}

We applied our motif cumulant based theory, in particular the relation between motifs and the response time constant, to recent network data on whole-brain mesoscale connectivity between 213 brain regions of the mouse brain (Fig.~\ref{fig:mouse_brain}A). The network is complex, as described by significant motif cumulants of many orders (Fig.~\ref{fig:mouse_brain}A). 

Our first result is that the motif structure of the brain-wide network extends the time over which the network remembers ``sensory'' input signals -- specifically, those passed into the network via the sensory thalamic nuclei (the brain areas that relay sensory input to cortical areas, Fig.~\ref{fig:mouse_brain}C).   We also find that the underlying motifs introduce multiple timescales into this process (the nonlinear decay on log-y plot of impulse response, Fig.~\ref{fig:mouse_brain}C-inset). Both of these effects largely vanish when we perform a node-degree preserving shuffle on the network (Fig.~\ref{fig:mouse_brain}C-inset). This shows these effects come specifically from higher order chain motifs which the shuffling destroys (in particular, the second order diverging and converging motifs (Fig. \ref{fig:example_motifs}A) are preserved, see Appendix \ref{sec:degree_shuffle}). The matching between the shuffled networks and the result of keeping only first order motifs $\tilde{\kappa}_1$ (Fig.~\ref{fig:mouse_brain}C-inset; see also Fig. \ref{fig:mouse_cortical_input_shuffle} for another example) also confirms our theory (Theorem \ref{TH:uniform_motif_multi}) in a real world network. 

The motif cumulant analysis also reveals interesting differences between how the mouse brain network responds to such sensory inputs vs. how it responds to ``top down" signaling from higher level brain areas that may be central to functions such as decision making.  While the exact list and function of such higher level areas in the mouse brain is the subject of ongoing research, we consider a proxy for higher-order inputs by passing input via cortical areas that are generally considered to be not immediately related to sensory or motor signals (see Appendix~\ref{sec:mouse_brain_area_list} for the list of areas we use).  We read out the response from all brain areas, precisely as for the sensory input case of Fig.~\ref{fig:mouse_brain}C. Comparing the network responses with sensory thalamic vs. higher-order cortical inputs, we find the following intriguing result:  While in  both cases the motif structures extend the time constant of the network response, higher-order cortical inputs lead to a significantly longer-timescale response (Fig.~\ref{fig:mouse_brain}D) -- consistent with what would be expected for, say, higher order signals that might slowly regulate brain or behavioral states. Moreover, longer chain motifs contribute significantly to the extension of time constant for higher-order cortical signaling, whereas the effect for the thalamic input mostly comes from the chains up to length 3 (Fig.~\ref{fig:mouse_brain}E,F; see also Fig. \ref{fig:mouse_cortical_input_shuffle}).

\subsection{Power grids and the \textit{C. elegans} neuronal network}
\begin{figure}[h!]
\begin{center}
\includegraphics[width=0.48\textwidth]{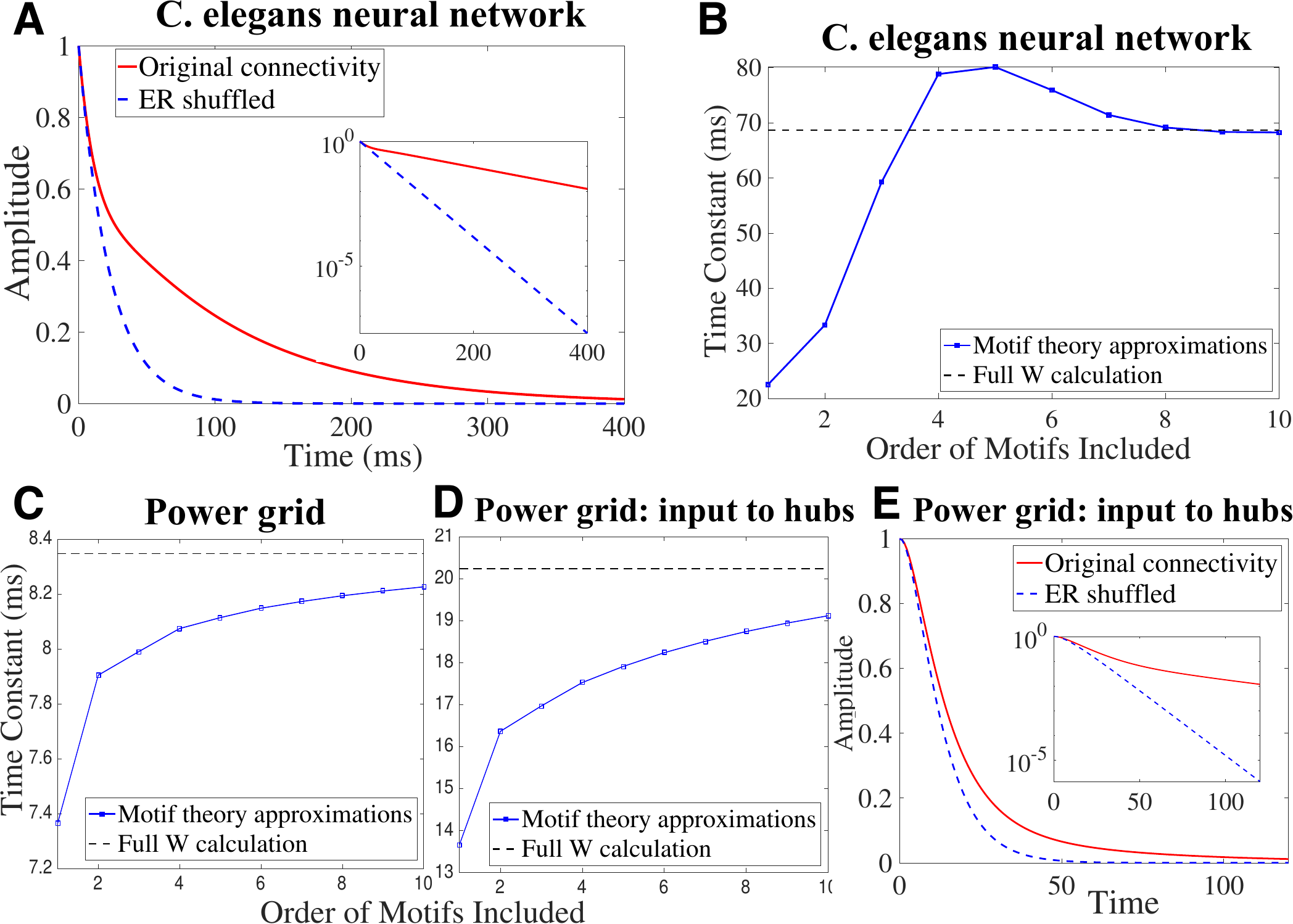}
\caption{{\bf Application to the \textit{C. elegans} neuronal network and a power grid network.}
{\bf A}. The \textit{C. elegans} network \cite{Watts:1998db} shows long timescale response (red solid line) that is much shortened when higher order motif cumulants are removed (blue dashed line). 
{\bf B}. Similar as Fig.~\ref{fig:mouse_brain} E,F; How motifs of various orders contribute to the extension of the time constant seen in A. 
For the power grid network of western states in US (C-E)\cite{Watts:1998db}, when the stimulus is sent to all the nodes, the motifs has only small effects on the impulse response ({\bf C}). In contrast, when input to only the hub nodes (4.2\% of nodes with the highest degrees), the impulse response has a larger time constant and the effect of higher order motifs become larger ({\bf D, E}). The original network with 4941 nodes is undirected. In the simulation, edges are made bi-directionals. 
}
\label{fig:other_real_networks}
\end{center}
\end{figure}

We next apply our results to two further real world networks of broad interest: the \textit{C. elegans} neuronal network and a power grid network \cite{Watts:1998db} (Fig.~\ref{fig:other_real_networks}). First, for the  \textit{C. elegans} network (with uniform global input and readout to all neurons), we find that -- despite differences in the spatial scales (single neurons versus brain areas) and species (worm vs mouse)  when compared the mouse brain network studied above  -- we once again observe that contributions from motif cumulants both extend the time constant beyond that for a random shuffled network and result in multiple timescales in the network response(Fig.~\ref{fig:mouse_brain}, \ref{fig:other_real_networks}).   

For the power grid network receiving global input (to all nodes), we find that motifs (beyond first order) only have relative small effect on the response function (Fig.~\ref{fig:other_real_networks}C). However, if we instead deliver input selectively to only the high degree hub nodes in the network, the network response now has a much larger time time constant, and the contribution from higher order motifs becomes more pronounced (Fig.~\ref{fig:other_real_networks}D,E). This both suggests -- explains via a tractable set of connectivity features -- a distinct functional role for the hub nodes in long-timescale signaling across the entire power grid.

\section*{Summary}

There is vast interest in relating network structure and dynamics across many fields.  Recent advances have shown how highly local (and therefore easily quantifiable) connectivity features -- or {\it network motifs} -- predict global levels of synchrony in the intrinsic dynamics generated autonomously in networks~\citep{Hu:2013vh,Hu:2014gs}.  In this paper, we show that a similar approach bears fruit in predicting the network response to external, temporal stimuli.  In particular, we study the signal filtering property of a recurrently connected network of LTI units, or equivalently applicable to, linearly interacting point process networks. We showed that the network transfer function is exactly determined by the chain motif cumulants of different lengths (Eq.~\eqref{E:transfer_motifs}). These measure the extent of {\it overrepresented} consecutive paths through the network, compared with what is expected from a hierarchy of lower-order graph statistics.  Importantly, only a few lower order motif cumulants are often needed to form accurate predictions of the network transfer function. Our theory thus provides a new way of inferring a basic \emph{global} functional property of a network based on partial, \emph{local} observations of its connectivity. Our approach is complementary to spectral graph theory~\cite{chung1997spectral}:  indeed, the eigenvalues of a graph and motifs can be directly related~\cite{preciado2013moment}.

For some networks, however, we showed a limitation of our first theory (Theorem~\ref{TH:uniform_motif}).  The approximations made by retaining lower order motif cumulant terms in Eq.~\eqref{E:transfer_motifs} start to deteriorate and may lead to an unstable approximation of the network transfer function, even though the true transfer function is stable and well defined. To resolve this issue, we developed an improved version of motif cumulant theory and a new corresponding formula relating these motif cumulants to the network transfer function.  This can be thought of as ``correcting" for heterogeneous degrees among the nodes (Section~\ref{S:degree_corrected_motif}). This degree-corrected theory improves the accuracy of motif-based predictions extending them to regimes where the original theory may not apply. 

The network time constant, which measures for how long past signals influence the future network response, is closely related to the network transfer function we study here. We explicitly link the time constant to the motif cumulants (Theorem~\ref{TH:time_constant}), and show across many networks that the presence of higher order motifs can lead to a large change of the time constant that would not be predicted based on the first order motif cumulant (connection probability) alone (Fig.~\ref{fig:time_constant_scatter}).

Beyond the idealized case where signals read into and out of networks via uniform weights, we considered two types of extensions. We generalized our theory to allow feeding input into or reading output out of a subset of nodes based on their population identity (Theorem~\ref{TH:uniform_motif_multi}). The connectivity statistics (motif cumulants) and dynamics of each node may also differ from one population to the next. This extension allows for broader applications of our theory and for a greater variety and richness of network transfer function to be predicted for a given network. 

Another important extension to our theory begins by considering the robustness of our theory to noise in the input and readout weights. We prove that large networks are indeed robust to the independent perturbations in these weights (Theorem~\ref{Th:converge}). Interestingly, however, if the fluctuations of weights are correlated from node to node, the resulting network transfer function will have an additional term that depends on the {\it cycle} (and chain) motif cumulants in the network (Eq.~\eqref{E:trace_resum}). How might such correlations -- and resultant changes in the network transfer properties -- arise?  One possible mechanism in the context of neural networks is through synaptic plasticity or learning processes. This points to interesting directions for future study where the input and output weights are outcomes of a learning process, during which the recurrent network is trained to achieve an input-output function~\citep{Sussillo2009,buonomano2009state}. To describe more powerful functionality, future work will also need to incorporate nonlinear dynamics at each node, over and above the linearization taken here.  We hope that the present work opens the door to such extensions, likely by providing the first terms in an expansion across orders of nonlinearity.

The present work, at the interface of network science, statistical physics, and systems engineering, has clear practical and scientific implications.  For one, localized connectivity, quantified through motifs, is relatively easy to sample and compare among systems~\cite{Milo2002science,Song:2005jy,Perin:2011cu}. Even more exciting, in neural systems this localized connectivity is under the control of learning, plasticity and adaptation mechanisms~\cite{Gerstner:2002us,Ocker2015}. Thus, our work may inspire new analyses of the natural learning and adaption of network function.

\section*{Acknowledgements}
We thank the anonymous referees for suggestions and critiques that have substantially strengthened the paper.  This work was supported by an AFOSR grant FA9550-09-0174 to JNK, and by a NIH training grant 5T90DA03243602, as well as NSF grant DMS-1122106 and a Simons Fellowship in Mathematics to ESB.  
K. Josic and M. Mesbahi provided very helpful insights and comments. We thank J. Harris for the list of associational cortical areas.
We thank the Allen Institute founders, Paul G. Allen and Jody Allen, for their vision, encouragement and support.

\section*{Appendix}
\begin{appendices}

\section{Use LTI to model linearly interacting point process}
\label{sec:point_process}
Here we show that a similar set of equations with LTI units as Eq.~\ref{E:node} can be used to model interacting point process~\cite{Hawkes:1971wh, Pernice:2011fr}. Let $\lambda_i(t)$ be the (stochastic) instantaneous firing rate of neuron $i$. Spike trains (point process) $S_i(t)$ are generated by an inhomogeneous Poisson process according to the rate $\lambda_i(t)$. We assume that $\lambda_i(t)$ is always or for most of the time positive, so that it serves as a legitimate Poisson rate. Neurons interact with each other through the spike trains filtered by a nodal kernel $h(t)$
\beq
\lambda_i(t)=\lambda_0 + \int_{-\infty}^t  h(t-t^\prime) \left(\sum_j W_{ij} S_j (t^\prime) +B_i u(t^\prime) \right)dt^\prime.
\label{E:Hawkes}
\eeq
Here $\lambda_0$ is a constant baseline firing rate. In absence of signal $u(t)$, the steady state firing rate $\bar{\lambda}_i$ satisfies
\[
\bar{\lambda}_i=\lambda_0+h_0\sum_j W_{ij} \bar{\lambda}_j,
\]
where $h_0=\int_0^\infty h(t)dt$.

Now we consider how the average firing rate changes over time when there is a time dependent signal $u(t)$. Let $\Delta \lambda_i(t)=\EVs{\lambda_i(t)}-\bar{\lambda}_i$, where the expectation is taken across trials while fixing the signal $u(t)$. For example, $\EVs{S_i(t)}=\EVs{\lambda_i(t)}$. Taking expectation over Eq.~\eqref{E:Hawkes}, we have
\beq
\Delta \lambda_i(t)=\int_{-\infty}^t  h(t-t^\prime) \left(\sum_j W_{ij} \Delta \lambda_j (t^\prime) +B_i u(t^\prime) \right)dt^\prime.
\label{E:Hawkes_E}
\eeq
This is the same set of equations for a network of linear time invariant (LTI) nodes.

\section{ Proof of a random matrix property Lemma~\ref{TH:WTh_tnorm}}
\label{S:proof_ER_kappa}

\thmWThtnorm*

\begin{proof}
We split the norm into two terms,
\beqr
\nonumber
&&\tnorm{W\Theta} = \tnorm{W-pNee^\tsp+pNee^\tsp-Wee^\tsp} \\
&&\leq \tnorm{W-pNee^\tsp}+ \tnorm{pNee^\tsp-Wee^\tsp} 
\label{E:split_WT}
\eeqr
The bound of the first term is a typical result in random matrix theory, for which we will rely heavily on the reference~\citep{vershynin_2012}. Note that $W-pNee^\tsp$ has Bernoulli	distributed i.i.d. entries. According to Lemma 5.24 in \citep{vershynin_2012}, the rows of this matrix are independent sub-gaussian isotropic random vectors (Definitions 5.22 and 5.19 in \citep{vershynin_2012}) with sub-gaussian norm bounded by some absolute constant $K$. We can then apply Theorem 5.39 in \citep{vershynin_2012} about matrices with sub-gaussian rows to $W-pNee^\tsp$, with for example $t=\sqrt{N}$. The theorem shows 
\beq
\tnorm{W-pNee^\tsp}\leq C\sqrt{N}
\label{E:norm_1}
\eeq
is satisfied with probability of at least $1-2\exp(-cN)$. Here $C$ and $c$ are constants only depending on the sub-gaussian norm bound $K$ and thus are also absolute constants.  As $N\rightarrow \infty$, Eq.~\eqref{E:norm_1} holds with at least probability $1-2\exp(-cN)\rightarrow 1$.

For the second term in \eqref{E:split_WT}, let $\mu_i=1/N\sum_j W_{ij}$ and the vector $\Delta \mu=(\mu_1-p,\mu_2-p,\ldots,\mu_N-p)^\tsp$.  Then the relevant matrix can be written as
\[
Wee^\tsp-pNee^\tsp=\sqrt{N} \Delta \mu e^\tsp,
\]
which is a rank-1 matrix. Therefore its two norm is  
\[
\tnorm{Wee^\tsp-pNee^\tsp}=\sqrt{N}\tnorm{\Delta \mu}\tnorm{e}=\sqrt{N}\tnorm{\Delta \mu}.
\]
The norm of the vector
\[
\tnorm{\Delta\mu}^2=\sum_i(\mu_i-p)^2
\]
is a sum of i.i.d variables $(\mu_i-p)^2$. It is straightforward to calculate the mean and variance of this norm, and we will then use the Chebyshev inequality to give a bound.

For ease of notation, let $W_j=W_{ij}$ (since $i$ is fixed) and $\Delta W_j=W_j-p$.  We have
\[
\EVs{(\mu_i-p)^2}=\var(\mu_i)=\frac{1}{N^2}\sum_j \var(W_{ij})=\frac{p(1-p)}{N},
\]
\noindent and
\beqrn
&&\var((\mu_i-p)^2)= \EVs{(\mu_i-p)^4}-\frac{p^2(1-p)^2}{N^2}\\
&=& \EVs{(\frac{1}{N}\sum_j \Delta W_j)^4}-\frac{p^2(1-p)^2}{N^2}\\
&=& \frac{1}{N^4}\mathbf{E} \left\{\sum_j \Delta W^4_{j}+4\sum_{j\neq k}\Delta W^3_{j}\Delta W_{k} \right. \\
&& 3\sum_{j\neq k} \Delta W^2_j \Delta W^2_k
+6\sum_{j\neq k\neq l} \Delta W^2_j \Delta W_k \Delta W_l\\
& &\left.+\sum_{j\neq k\neq l\neq m}\Delta W_j \Delta W_k \Delta W_l \Delta W_m \right \}
-\frac{p^2(1-p)^2}{N^2}\\
&=& \frac{m_4}{N^3}+0+\frac{3(N-1)p^2(1-p)^2}{N^3}+0-\frac{p^2(1-p)^2}{N^2}\\
&=& \frac{m_4}{N^3} +\frac{(2N-3)p^2(1-p)^2}{N^3} \leq \frac{3 p^2(1-p)^2}{N^2} \\
&& \text{  for large enough $N$ (e.g. for $N\ge \frac{m_4}{p^2(1-p)^2}$).}
\eeqrn
Here $m_4=p-4p^2+6p^3-3p^4$ is a constant.

Applying the Chebyshev inequality to the variable $\tnorm{\Delta \mu}^2$, which has mean $p(1-p)$ and variance less than $\frac{3p^2(1-p)^2}{N}$, we have
\beqrn
&&P\left(\tnorm{\Delta \mu}^2 >p(1-p)+ \sqrt{N}\frac{\sqrt{3}p(1-p) }{\sqrt{N}} \right. \\ 
&& \left.=(1+\sqrt{3})p(1-p) \right) <\frac{1}{N}.
\eeqrn
This shows that the probability
\beqr
\nonumber
&&P\left(\tnorm{Wee^\tsp-pNee^\tsp}=\sqrt{N}\tnorm{\Delta \mu} \right.\\
&& \left. \leq \sqrt{(1+\sqrt{3})p(1-p) } \sqrt{N} \right)\ge 1-\frac{1}{N},
\label{E:norm_2}
\eeqr
so that this probability goes to 1 as $N\rightarrow \infty$.

Finally, combining the results from \eqref{E:norm_1} and \eqref{E:norm_2} with Eq.~\eqref{E:split_WT} and using the ``union bound"  ($P(A \cup B) \le P(A) + P(B)$), we have 
\[
\tnorm{W\Theta} \leq \tnorm{W-pNee^\tsp}+\tnorm{pNee^\tsp-Wee^\tsp}\leq C \sqrt{N}
\]
with probability approaching 1 as $N\rightarrow \infty$, for some absolute constant $C$.
 
\end{proof}

\section{Equivalence in stability for global feedback and mean-shifted networks}
\label{sec:stability_mean_shift_W}
To develop the corresponding stability condition, we imagine that there is an additional, $N+1^{st}$ node that functions as the global feedback. The node will have a filter that multiplies its input by a constant, and receives input from, and provides output to, all nodes in the original network. The stability of this $N+1$ node system is determined by studying for what values of $s\in\mathbb{C}$ the following matrix is invertible~\citep{Ogata:2010tz}
\begin{align*}
&I_{(N+1)\times(N+1)} -\left[ \barr{cc} h(s) W & h w_0 e \\ e^\tsp & 0  \earr \right]\\
&=\left[ \barr{cc}  1-h W & -h w_0 e \\   e^\tsp & 1  \earr \right].
\end{align*}
Using the Schur complement, this matrix is invertible if and only if $I-h(s)(W+\frac{w_0}{N} e e^\tsp)$ is invertible. This is precisely the stability condition for the network with all entries being shifted by $\frac{w_0}{N}$, and shows the equivalence between global feedback and adjusting $\kappa_1$ in terms of stability.

\section{Explaining effects of $\kappa_n$ based on poles of $G(s)$}
\label{S:pole_kappa}

Given the form of Eq.~\eqref{E:transfer_motifs}, the poles of $G(s)$ can be determined by studying the roots of the denominator. Assuming that the first $m$ of $\kappa_n$ are significant while $\kappa_{n>m}\approx 0$, the set of poles of $G(s)$ is 
\begin{align*}
&h^{\text{inv}}(z),~z\in \text{root}\left\{P(z) = \right. \\
&\left.  N^m\kappa_m z^m+ 
N^{m-1}\kappa_{m-1} z^{m-1}+\cdots
+N\kappa_1 -1\right\}.
\end{align*}
Here $h^{\text{inv}}(\cdot)$ is the (complex) inverse function of $h(s)$. For the two filters $h_{\exp} (s),\;h_{\cos}(s)$ we use (Appendix \ref{sec:node_filter}), $h^{\text{inv}}(s)$ can be readily characterized analytically.  Moreover, if we idealize the networks above so that only $\kappa_1$ and/or $\kappa_2$ are nonzero, $P(z)$ is a first or second order polynomial, which is readily solvable.  Taking these facts together, one can explain the qualitative changes seen in the network transfer function $G(s)$ when $\kappa_1$ and $\kappa_2$ are varied.  This includes the slower temporal decay seen with positive $\kappa_1$ in the case of $h_{\exp}(s)$, the split of the resonant peak with negative $\kappa_2$ in the case of $h_{\cos}(s)$, and so on. Details of the analysis are given below.

The number of poles of $G(s)$ is determined by the number of roots of $P(z)$ and $\# h^{\text{inv}}(z)$. Here $\# h^{\text{inv}}(z)$ is the number of pre-images of $z$ under the mapping $h(s)$. It can be shown that $\# h^{\text{inv}}_{\exp}(z)\equiv 1$ for all $z\in \mathbb{C}$ and $\# h^{\text{inv}}_{\cos}(z) \equiv 2$ for all $z\in \mathbb{C}\backslash \{ -\frac{1}{2\nu}, \frac{1}{2\nu}\}$. Therefore, when using these node transfer functions, for most cases, the number of poles for the network transfer will be determined by the number of roots of $P(z)$.  This number is the degree $m$ by the fundamental theorem of algebra. In particular, we conclude that the numbers of poles of $G(s)$ are 1 for $h_{\exp}$ and 2 for $h_{\cos}$ in the case of $\kappa_1\neq 0,~\kappa_{n\ge 2}=0$; and are 2 for $h_{\exp}$ and 4 for $h_{\cos}$ in the case of $\kappa_2\neq 0,~\kappa_{n\ge 3}=0$. This difference in the number of poles as the network connectivity changes reflects the qualitative changes that we see in the network transfer functions $G(s)$ (Fig.~\ref{F:change_kappa2_exp} and \ref{F:change_kappa2_cos}).  

To make this more precise, we study the location of the poles. For the exponential filter $h_{\exp}$, there is one (simple) pole $-\alpha$ on the real line. The decaying-oscillatory filter $h_{\cos}$ has a pair of complex conjugate poles $-\alpha\pm \nu i$. In general, a pair of complex conjugate poles will give rise to a peak in the frequency domain and real poles correspond to exponential decay.  As we have seen in these two examples, the magnitude of the real part of the poles determines the speed of decay in time domain, or the width of the frequency peak. A last fact that we will use is the symmetry of $h(s)$ with respect to complex conjugate, that is $\overline{h(s)}=h(\bar{s})$. 

For the case of $\kappa_1\neq 0,~\kappa_{n\ge 2}=0$, $P(z)$ always has one real root $\frac{1}{N\kappa_1}$. By the symmetry under conjugacy, this real root is mapped to one real pole under $h^{\text{inv}}_{\exp}(z)$ and to a pair of conjugate poles under $h^{\text{inv}}_{\cos}(z)$. As $\kappa_1$ increases (decreases), the real parts of those poles, $N\kappa_1-\alpha$ and $N\kappa_1/2-\alpha$ respectively for the two filters, increase (decrease) and result in the changes of time constant ($h_{\exp}(s)$) or peak width ($h_{\cos}(s)$).

For the other case where $\kappa_2\neq 0$, $\kappa_1=\kappa_{n\ge 3}=0$, $P(z)$ can either have two real roots or a pair of complex conjugate roots, depending on the sign of $\kappa_2$. When $\kappa_2>0$, each of the real roots is similarly mapped under $h^{\text{inv}}(z)$ as in the previous case. The poles corresponding to one of the two real roots dominates the effect on the time constant or peak width. When $\kappa_2<0$, the complex root pair is mapped to two complex poles under $h^{\text{inv}}_{\exp}(z)$, which generate the observed oscillation in the impulse response. Under $h^{\text{inv}}_{\cos}(z)$, each complex root is mapped to two non-conjugate complex poles with different imaginary parts, while the image of the other root is exactly the conjugate of these poles. All together, the complex root pair of $P(z)$ are mapped into four poles, occurring as two conjugate pairs. The fact that these pole pairs have different imaginary parts explains the observation of two distinct frequency peaks.

The arguments above show how the roots of $P(z)$ determine qualitative properties of the network transfer function $G(s)$.  This is interesting, because $P(z)$ is determined completely by the network's motif statistics, independent of the nodal dynamics $h(s)$.  For example, in Fig.~\ref{F:change_kappa2_exp} and \ref{F:change_kappa2_cos}, as $\kappa_2$ becomes negative, the roots of $P(z)$ switch from two real ones into a complex pair. Correspondingly, $G(s)$ seems to undergo a type of ``bifurcation" in the case of both $h(s)$ functions. We suggest that although the specific changes in $G(s)$ depend on details of $h(s)$, the onset of the transition is often determined by alone $P(z)$, which is also a form of ``generating function" for motif cumulants.

\section{Proof of the convergence of $G(s)$ under independent random input output weights}
\label{sec:proof_converge_G}
\thmconverge*
\begin{proof}
First, we will show that the convergence $\kappa_n \rightarrow \kappa_n^\infty$ along with \eqref{E:W_norm_condition} implies the convergence of $\EVs{G(s)}\rightarrow G^{\infty}(s)$. Note that 
\[
\tnorm{\Theta W\Theta}\leq \tnorm{\Theta} \tnorm{W} \tnorm{\Theta} =\tnorm{W}.
\]
Therefore, by \eqref{E:W_norm_condition},
\[
\frac{1}{N}\abs{h(s)}\tnorm{\Theta W\Theta}\leq \frac{1}{N}\abs{h(s)}\tnorm{ W} \leq 1-\delta.
\]
This inequality can be used with the matrix expression for $\kappa_n$, Eq.~\eqref{E:kappa_matrix_ch}, to show that $\abs{h^n(s)\kappa_n} \leq (1-\delta)^n$. This geometric bound (in $n$) guarantees that $\sum_{n=1}^{\infty} \abs{h^n(s)\kappa_n}$ is bounded independent of $N$ and $W$, and $\sum_{n=1}^{\infty} h^n(s)\kappa_n$ converges absolutely to $\sum_{n=1}^{\infty} h^n(s)\kappa^\infty_n<\infty$ (by dominant convergence). The above leads to 
\begin{align*}
&\EVs{G(s)}=\left(1-\sum_{n=1}^{\infty} h^n(s)\kappa_n\right)^{-1}h(s)\\ 
&\rightarrow \left(1-\sum_{n=1}^{\infty} h^n(s)\kappa^\infty_n \right)^{-1}h(s) =G^\infty(s).
\end{align*}

Furthermore, the convergence is uniform in $s$ as we show below. For any $\epsilon>0$, there exists an integer $n_1$ such that $\frac{(1-\delta)^{n_1+1}}{\delta}<\frac{\epsilon}{4}$. For each of $1\leq n \leq n_1$, since $\kappa_n\rightarrow \kappa_n^\infty$, there exists an integer $N_n$ such that for any $N>N_n$, 
\[
\abs{\kappa_n-\kappa_n^\infty}\leq \frac{\epsilon}{2 n_1 (\max_s \abs{h(s)})^n}.
\]
Let $N_\epsilon=\max_{1\leq n \leq n_1} N_n$. For any $N>N_\epsilon$,
\beqrn
&&\abs{\sum_{n=1}^{\infty} h^n(s)\kappa_n -\sum_{n=1}^{\infty} h^n(s)\kappa^\infty_n} \leq
\abs{\sum_{n=1}^{n_1} h^n(s) (\kappa_n-\kappa_n^\infty)}\\
&&+\abs{\sum_{n=n_1+1}^{\infty} h^n(s) \kappa_n }+\abs{\sum_{n=n_1+1}^{\infty} h^n(s) \kappa_n^\infty }\\
&&\leq \sum_{n=1}^{n_1}\frac{\epsilon}{2 n_1}+2\sum_{n=n_1+1}^\infty (1-\delta)^n\\
&&=\frac{\epsilon}{2}+\frac{2(1-\delta)^{n_1+1}}{\delta}=\epsilon.
\eeqrn
This shows that
\beq
\label{E:series_converge_uniform}
\sum_{n=1}^{\infty} h^n(s)\kappa_n \rightarrow \sum_{n=1}^{\infty} h^n(s)\kappa^\infty_n 
\quad \text{uniformly in $s$}.
\eeq
Using the following inequality
\begin{align*}
&\abs{\EVs{G(s)}-G^\infty(s)}\leq \max_{s} \abs{h(s)} \cdot \\ 
&\left(
\left(1-\sum_{n=1}^{\infty} h^n(s)\kappa_n\right)^{-1} -\left(1-\sum_{n=1}^{\infty} h^n(s)\kappa^\infty_n\right)^{-1} 
 \right),
\end{align*}
and composing the limit of Eq.~\eqref{E:series_converge_uniform} with function $(1-x)^{-1}$,  we conclude the uniform convergence of $\EVs{G(s)}$ to $G^\infty(s)$ in $s$.

We will now use the Chebyshev's inequality to show the convergence of $G(s)$ to $\EVs{G(s)}$ based on calculating the variance of $G(s)$. Let $P=(I-h(s)W)^{-1}$, we have
\beqrn
&&\EVs{\abs{G(s)}^2}:=\EVs{C^\tsp P B C^\tsp \bar{P} B} \abs{h(s)}^2 \\
&=&\EVs{C^\tsp P B B^\tsp P^* C}\abs{h(s)}^2\\  
&=& \EVs{\tr( P B B^\tsp P^* C C^\tsp)} \abs{h(s)}^2\\
&=& \tr(P\EVs{ B B^\tsp}P^*\EVs{C C^\tsp})\abs{h(s)}^2\\
&=&\tr(P( e e^\tsp +\sigma^2 I)P^*(e e^\tsp +\sigma^2 I))\abs{h(s)}^2.
\eeqrn
Here $\bar{P}$ is the entry-wise complex conjugate and $P^*=\bar{P}^\tsp$. In the fourth equality, we have used the property that $B$, $C$ are independent. 

Combining this with the expression for the mean (Eq.~\eqref{E:E_Gs_ind}) we have
\begin{eqnarray}
\nonumber
&&\Var(G(s)):=\EVs{\abs{G}^2}-\abs{\EVs{G}}^2\\
\nonumber
&=& \sigma^4 \tr(PP^*) \abs{h(s)}^2+\sigma^2 
(e^\tsp PP^*e+e^\tsp P^* Pe)\abs{h(s)}^2\\
\nonumber
&=& \frac{1}{N^2} \tr(PP^*) \abs{h(s)}^2\\&&+\frac{1}{N} 
(e^\tsp PP^*e+e^\tsp P^* Pe)\abs{h(s)}^2.
\label{E:var_rand_ind_weight}
\end{eqnarray}
Given the norm condition on $W$, we have
\[
\tnorm{P}=\tnorm{P^*} \leq \frac{1}{1-\frac{1}{N}\tnorm{h(s)W}}\leq \frac{1}{\delta},
\]
which is a constant bound independent of $N$ and $W$. Using this,
\[
\tr(PP^*)=\tr(P^*P) \leq N \tnorm{P}^2 \leq \frac{N}{\delta^2},
\]
\[
\abs{e^\tsp PP^*e} \leq \tnorm{e^\tsp}\tnorm{P}\tnorm{P^*} \tnorm{e} \leq \frac{1}{\delta^2},
\]
similarly 
\[
\abs{e^\tsp PP^*e} \leq \frac{1}{\delta^2}.
\]
Together, we have
\[
\Var(G(s)) \leq \frac{3 \abs{h(s)}^2}{N\delta^2} \rightarrow 0, \;\; \text{as } N\rightarrow \infty.
\]
Chebyshev's inequality then ensures the convergence of $G(s)$ in probability.
\end{proof}

\section{Comparing motif magnitude across different orders}
\label{sec:motif_magnitude_scaling}
When comparing the magnitude of motifs across different orders, the higher order motif cumulants are usually much smaller because they contain more edges and are more rare to occur. To compensate this intrinsic difference of scales, for binary networks where all connections are of the same strength (e.g. in Fig. \ref{fig:ladder_diagram}A), we compute the relative magnitude $\kappa_n / p^n$, where $p$ is the connection probability. 

For weighted networks where the connection strengths can be any real number, the motifs cumulants are weighted, which leads to an ambiguity in interpreting the magnitude of motifs. If we multiply $W$ with a positive constant $\gamma$, its ``graphical" properties are unchanged, but the motif cumulants will not be scaled as $\kappa_n \gamma^n$. To address this, for weighted networks (e.g. in Fig. 17B), we scale $W$ such that its spectral radius is $0.9$ and calculate the motif cumulants under such a scaling.

\section{Estimating motif cumulants $\kappa_n$ by local sampling of connectivity}
\label{S:estimating_motif_sampling}

The motif cumulant $\kappa_n$ can be estimated by randomly sampling $n+1$ nodes $i_1,\ldots,i_{n+1}$  in the network (equivalent to having access to a random $(n+1) \times (n+1)$ diagonal block of the full connectivity matrix $W$). The procedure is as follows.  For each of the samples, record whether there is a length-$n$ chain $i_1\rightarrow i_2 \rightarrow \cdots \rightarrow i_{n+1}$. After a sufficient number of such samples, the proportion of samples having the chain structure gives an estimate of the motif moment $\mu_n$. For weighted motifs, we will replace each motif count by the product of connection weights it contains. Once the estimates of all motif moments $\mu_{n^\prime \leq n}$ are gathered, we can use the decomposition relation Eq.~\eqref{E:mu_kappa} between motif moments and cumulants to calculate $\kappa_n$.

The following result shows that the sampling method given above leads to an unbiased estimator of $\mu_n$. Note that the same proof can be used to establish the result for the case of weighted motifs. In the large sample limit when the estimates for $\mu_n$ converge, $\kappa_n$ can be correctly estimated.
\begin{lemma}
Let $0\leq i_1,\ldots,i_{n+1} \leq N$ be ordered, randomly sampled indices, allowing duplications, of a network of size $N$ with adjacency matrix $W$. We have
\[
P(\text{there is a chain }i_1\rightarrow i_2 \rightarrow \cdots \rightarrow i_{n+1}) = \mu_n,
\]
where $\mu_n$ is the motif moment of the network.
\end{lemma}

\begin{proof}
Let $ \mathbf{1}_{i_1\rightarrow i_2 \rightarrow \cdots \rightarrow i_{n+1}}$ be the indicator random variable for whether there is a length $n$ chain $i_1\rightarrow i_2 \rightarrow \cdots \rightarrow i_{n+1}$. Note that for each possible index sample $i_1,\ldots,i_{n+1}$, its probability of being chosen in the above sampling scheme is the same, and is equal to $1/N^{n+1}$. We have
\begin{align*}
& P(\text{there is a chain }i_1\rightarrow i_2 \rightarrow \cdots \rightarrow i_{n+1})\\
&=\mathbf{E}  \;\mathbf{1}_{i_1\rightarrow i_2 \rightarrow \cdots \rightarrow i_{n+1}}\\
&= \mathbf{E} \;W_{i_{n+1}, i_{n}} \cdots W_{i_3, i_2} W_{i_2, i_1}\\
&=\sum_{i_1,\ldots,i_{n+1} = 1}^N  \frac{1}{N^{n+1}} W_{i_{n+1}, i_{n}} \cdots W_{i_3, i_2} W_{i_2, i_1} \\
&= \frac{1}{N^{n+1}}  e^T W^n e
 = \mu_n.
 \end{align*}
Here $e=(1,\ldots,1)^T$ is the uniform $N$-vector. We have used the definition of motif moments (Sec.~\ref{S:main_result}) in the last equality.
\end{proof}

We make a number of additional remarks about the random sampling method. First, and very importantly for practical applications, the method only requires sampling the network \emph{locally} at one time and is thus compatible with how connectivity motifs are measured in many real world applications such as neuronal networks \citep{Song:2005jy,Perin:2011cu}. Moreover, to use sampling data more efficiently, each time when we sample $n+1$ nodes, we can also  re-sample and sub-sample from these $n+1$ nodes (allowing duplicated indices). The resampling generates more samples ($(n+1)^{n+1}$ to be precise) for the estimation of $\mu_n$. The contribution of these resampled motifs to the estimator of $\mu_n$ can be directly calculated as $\frac{1}{(n+1)^{n+1}} \hat{W}^{n}$, where $\hat{W}$ is the diagonal block of $W$ corresponding to indices $i_1,\ldots,i_{n+1}$. The subsamples of $n^\prime +1 \leq n+1$ indices from $i_1,\ldots,i_{n+1}$ can be used to estimate $\mu_{n^\prime}$. Similarly the contribution of these subsamplings can be directly calculated as $\frac{1}{(n+1)^{n^\prime+1}} \hat{W}^{n^\prime}$. We note that the random samples are in general not independent so the variance of the estimator cannot not be simply derived based on the number of samples. However, we intuitively expect the correlations between the samples to be small when the network size is large compared to the size of motifs being estimated. 

The level of fluctuation in the estimators of motifs will also depend on the level of heterogeneity in the connectivity. We expect the estimators to converge faster in more homogeneous networks; and for strongly heterogeneous networks, the multiple population theory (Sec.~\ref{S:multi_pop}) is probably more appropriate (see also \citep{Hu:2014gs} on how to identify the populations having distinct connectivity statistics). A more complete discussion of the convergence of these estimators of motifs as well as exploration of more sophisticated estimation methods is an important topic but is beyond the scope of this paper.

\section{Shuffling connections to remove higher order motifs}
\label{sec:degree_shuffle}
Here we describe details on the degree-preserving shuffling used in Fig.~\ref{fig:mouse_brain}C to isolate the impact of network structure on network response. This procedure will result in a random graph with an in-degree and out-degree distribution (and connection weight distribution) identical to the original network, but with the sources and targets of each node redrawn independently. This can be seen by noting the row-sum and column-sum (i.e. the in-degree and out-degree of a node) of the original (left) matrix is the same as the final (right) matrix. Importantly, this means the second order converging and diverge motifs (Fig. \ref{fig:example_motifs}; see also the definition in Appendix \ref{S:gaussian}) are preserved because they can be expressed as variances of in- and out-degrees respectively \cite{Zhao:2011dv,Hu:2013vh}. 

On the other hand, the values in each row and column will be in a different order, resulting in independent in- and out-degree distributions. Thus, any remaining network structure is due to unequal weights and degrees, but not due to any special (i.e. nonrandom) configuration of connections beyond this. Consequently, all higher order chain motif cumulants but $\kappa_1$ are reduced to zero.

\begin{figure}[h!]
\begin{center}
\includegraphics[width=0.45\textwidth]{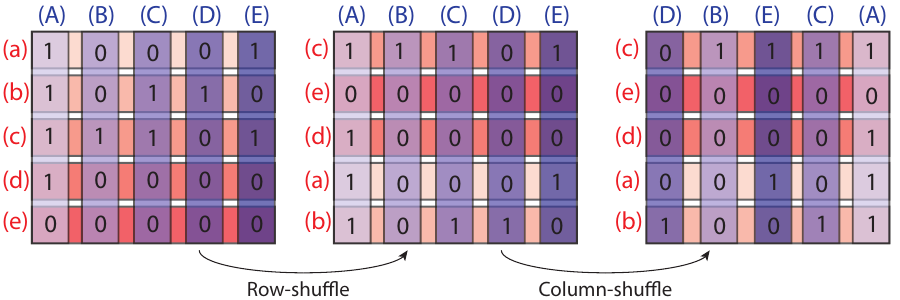}
\caption{{\bf Degree distribution preserving shuffling of network connectivity.}  The shuffling produces a random network by transforming the connectivity matrix for a network via a random row permutation followed by a random column permutation. The order of shuffling rows first and then columns (as depicted here) is arbitrary, and can be reversed.}
\label{fig:mouse_shuffle}
\end{center}
\end{figure}

\begin{figure}[h!]
\begin{center}
\includegraphics[width=0.3\textwidth]{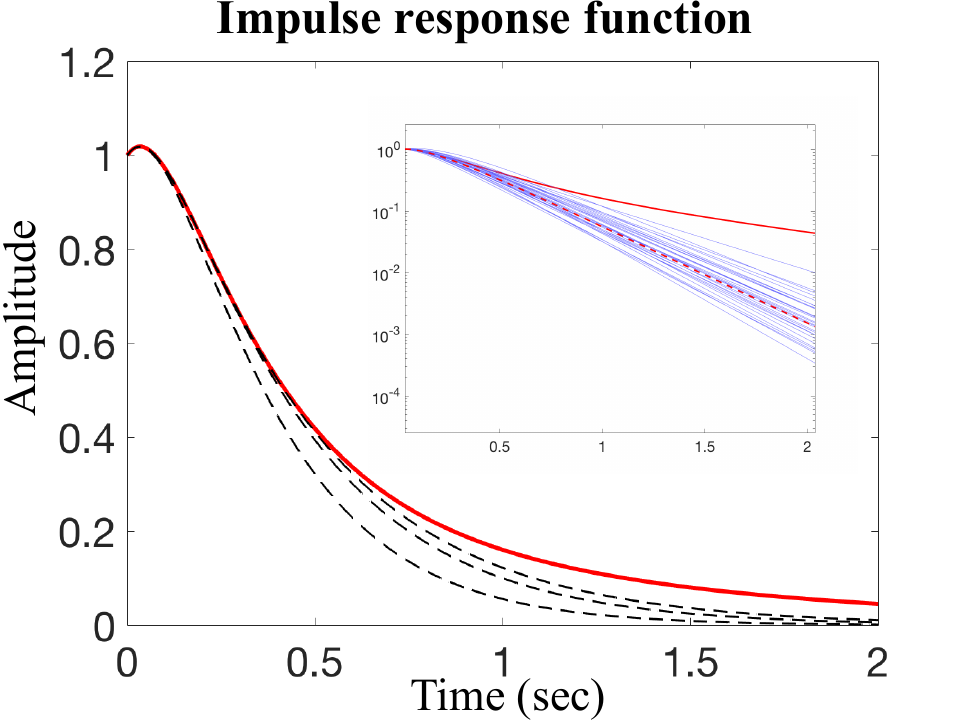}
\caption{{\bf Comparison of degree-preserving shuffle and the theory of truncating motifs.} Same as Fig.\ref{fig:mouse_brain}C but for the case of sending input to cortical areas.}
\label{fig:mouse_cortical_input_shuffle}
\end{center}
\end{figure}

In Fig.~\ref{fig:mouse_brain}C (and its counterpart for cortical input Fig. \ref{fig:mouse_cortical_input_shuffle}), we generate 100 samples of networks by independently shuffling the rows and columns of the connectivity matrix $W$ (Fig.~\ref{fig:mouse_shuffle}) of the mouse brain network. In these applications, because we send input to a subset of brain areas (sensory thalamic areas or cortical areas), we respect such a distinction in the shuffling. In particular, we shuffle over each of the four blocks of the connectivity matrix formed by two groups of areas: input-receiving areas and the rest. Therefore, the connectivity organization at the broad level of the groups is respected, while any higher order chain motif structures are removed.

\section{Network generation methods}
\label{S:network_generation}

We use two classes of random networks in our numerical examples, as described in detail below.  The first is the class of ``sparse" networks with the majority of entries in the connection matrix $W$ being 0; these are generated according to the \ER{} and the second order network (SONET) models.  The second class is dense networks, with most entries in $W$ being non-zero and taking continuous values; these networks are generated via Gaussian random matrices.  

The \ER{} network is generated by simply drawing each connection independently as a Bernoulli random variable with connection probability $p$.  We next give the details by which the other networks are generated.

\subsection{Generating Gaussian networks with chain, converging, and diverging motifs}
\label{S:gaussian}
We first consider a network with Gaussian distributed entries, $W_{ij}=a_i+b_j+c_{ij}$, where $a_i$, $b_j$ and $c_{ij}$ are all Gaussian variables with zero means.  Furthermore, assume that all of these variables are independent, {\it except} for pairs $(a_i,b_i)$, $i=1,\cdots,N$.  Then, it is easy to verify that $\cov(W_{ij},W_{jk})= \cov(a_j,b_j)$, $\cov(W_{ij},W_{ik})=\var(a_i)$ and $\cov(W_{ik},W_{jk})=\var(b_k)$. By considering the corresponding indices one sees that these covariances correspond to the excess probability, or cumulants, of length-two chain, converging and diverging motifs for large networks (assuming $a_i$, $b_i$, $c_{ij}$ and $(a_i,b_i)$ have identical distributions across $i,j$). 
\beqrn
&&\cov(W_{ij},W_{jk})=\kappa_{2},\\
&& \cov(W_{ij},W_{ik})= \kappa_{con},\quad \cov(W_{ik},W_{jk})= \kappa_{div}
\eeqrn
By adjusting the variance and covariance of $a_i, b_i$, we can therefore achieve various values of motif cumulants.  One can show that the resulting motif cumulants must satisfy the following inequality constraints
\[
\kappa_{con}+\kappa_{div}\leq \sigma^2,\quad  \abs{\kappa_2}\leq \sqrt{\kappa_{con}\kappa_{div}}
\]
Here $\sigma^2$ is the variance of the entries of $W$ (except for entries on the diagonal).

\subsection{Generating sparse complex networks with the SONET graph model}
\label{S:sonet}
We use the SONET model of random graphs, together with code provided by the authors of~\citep{Zhao:2011dv}, to generate sparse networks with different motif statistics. As an extension of the \ER{} model, the algorithm generates a $W$ with binary entries, with a given connection probability and approximately specified second order motif cumulants (for converging, chain, diverging and reciprocal connection motifs).

\subsection{Generating networks with different cycle motif cumulants $\kappa^c_2$}
\label{S:gaussian_cycle}
We achieve various values for the cycle motif cumulant $\kappa^c_2$ via another model of Gaussian random matrices, with an adjustable level of symmetry in the matrix entries.   First, we point out that $\kappa^c_2$ is directly related to the correlation coefficient $\rho_{reci}$ of entries in the connection matrix that correspond to reciprocal connections, such as $W_{ij}$ and $W_{ij}$. In particular, it can be shown that for large networks generated with Gaussian entries (assuming no correlations except for between reciprocal entries), $\kappa^c_2\approx \sigma^2 \rho_{reci}$, where $\sigma^2$ is the variance of $W_{ij}$.  The argument is as follows.  Below, we assume that the network size is large and replace the sum of large number of (nearly) independent variables by its expected value. One can show that 
\beqrn
&&\kappa_1=\frac{1}{N} e^\tsp W e=0, \\
&&\kappa_2=\frac{1}{N^2} e^\tsp W^2 e-\kappa_1^2= O(\frac{1}{N})-0 \rightarrow 0,\\
&&\kappa^c_2=\frac{1}{N^2}\tr(W\Theta W \Theta)=\frac{1}{N^2}\tr(W^2)-2\kappa_2-\kappa_1^2\\
&&=\rho_{reci} \sigma^2+O(\frac{1}{N})-0-0 \rightarrow \rho_{reci} \sigma^2.
\eeqrn

Finally, we can readily construct Gaussian random matrices with arbitrary levels of $\rho_{reci}$, while keeping all other correlations among entries of $W$ equal to 0. To do this, we generate a $W$ matrix as a weighted sum of a symmetric or anti-symmetric Gaussian matrix and an independent Gaussian matrix, with special treatment for the diagonal entries (to keep their variance the same as for other entries). This method allows to one achieve all possible range of $\rho_{reci}$ ($[-1,1]$).

\section{Additional details and parameters for numerical examples}
\label{S:numerical_detail}

\subsection{Nodal filters}
\label{sec:node_filter}
In the numerical examples, we set the node filter $h(s)$ (for all nodes in the network) to be one of two forms:  an exponential filter, or a decaying-oscillatory filter.  Specifically, we take:  
\beq
\label{E:filter_exp}
 h_{\exp}(t)=e^{-\alpha t } H(t),\quad  \mathcal{L} (h_{\exp}) (s)=\frac{1} {s+\alpha},  
\eeq
and
\begin{eqnarray}
\nonumber
&&h_{\cos}(t)=e^{-\alpha t} \cos(\nu t) H(t),\\
&&\mathcal{L} (h_{\cos}(s))=\frac{s+\alpha}{ (s+\alpha)^2+\nu^2} \; .
\label{E:filter_cos} 
\end{eqnarray}
here, $H(t)$ is the Heaviside function, and the Laplace transforms are given in parentheses.

When not stated otherwise, we set the parameters for $h(s)$ filters in Eq.~\eqref{E:filter_exp}  and \eqref{E:filter_cos} to be $\alpha=0.2$ and $\nu=2\pi/7$ with units of rad/s. We choose these values only for purpose of concreteness and plotting:  our results do not rely on these particular values, or on the units of these parameters. The parameters for the real world networks are set based on the context and described respectively for each case.

\subsection{Connection strength and stability condition for the network system}
For convenience, we describe the connection matrix up to a positive constant $a$ that determines the overall magnitude of the connection strength. For example, we may refer to $W$ as an \ER{} network with connection probability $p$, but the actual connection matrix is $\frac{1}{N}W$. This constant is not written explicitly, but is assumed to be absorbed into $W$. 

The constants in numerical examples are often chosen based on the largest possible connection strength that will keep the network system stable.  This largest value is determined by $W$ and $h(s)$, and can be efficiently computed using a semi-analytic method that we describe next.  The exact stability condition for any LTI system $x(s)=G(s)u(s)$ is that there is no pole on the right-half-plane of complex $s$ values. For our model, $G(s)=(I-ah(s)W)^{-1}$, and the condition on the poles can be translated into a condition based on the eigenvalues of $W$ and on a region in the complex plane defined by $h(s)$. 
The poles of $G(s)$ satisfies
\[
1=ah(s)\lambda_i, \text{  or   } \frac{1}{h(s)}=a\lambda_i
\]
where $\lambda_i$ is the eigenvalue of $W$. If we define a region in the complex plane
\[
\Omega:=\{1/h(z) |  \Re(z)>0 \},
\]
then the stability condition is equivalent to requiring that the point cloud of eigenvalues of $W$ scaled by $a$ does not fall in to $\Omega$.  

For the $h_{\exp}(s)$ and $h_{\cos}(s)$ functions we use (Appendix \ref{sec:node_filter}), this region $\Omega$ can be determined analytically. For $h_{\exp}(s)$, $1/h(z)=z+\alpha$, and $\Omega=\{z | \Re(z)>\alpha\}$. For $h_{\cos}(x)$ (while $\alpha<\nu$), the boundary of $\Omega$ is determined by the curve of $\{z+\nu^2/z | z=\alpha+x, x\in(-\infty,-\sqrt{\nu^2-\alpha^2})\cup(\sqrt{\nu^2-\alpha^2},\infty)\}$. In particular the boundary has a singular and right-most point at $2\alpha$. These characterizations make it easy to calculate the critical $a$ for stability.

\subsection{Parameters in Fig. \ref{fig:motif_effect_hexp}, \ref{fig:motif_effect_hcos}}

All network examples have 1000 nodes. The four networks on the axis of $\kappa_1$ (red, cyan, green and blue) are generated as \ER{} networks with connection probability ($\kappa_1$) 0.05, 0.1, 0.2 and 0.4. The networks with non-zero $\kappa_2$ (orange, pink) are generated as SONETs (Appendix \ref{S:network_generation}) with connection probability ($\kappa_1$) 0.1, and $\kappa_2=-0.6\times 10^{-2}$ and $0.6\times 10^{-2}$. In the bar plot of motif cumulants, we normalize $\kappa_{n\ge 2}$ as $\kappa_n/\kappa_1^n$.

\subsection{Parameters in Fig. \ref{fig:random_weights_convergence}}

In Fig. \ref{fig:random_weights_convergence}, we demonstrate the convergence described in Theorem \ref{Th:converge} with numerical examples. We generate two networks of size $N=100$ and $N=1000$ with Gaussian random $W$ (Appendix \ref{S:gaussian}) and node filter $h_{\exp}(s)$. The two networks are designed to have the same $\kappa_n$, so that, when their entries are scaled by $1/N$, the corresponding $G(s)$ functions are identical (with uniform weights $B, \, C=e$). Specifically the entries of $W$ have zero mean and variance $0.09$, are chosen to be correlated, so that $\kappa_2=-0.6\times 10^{-2}$ for both of the matrices. All other $\kappa_n$ are approximately 0 by construction.

The $G(s)$ under uniform weights are the red curves in Fig. \ref{fig:random_weights_convergence}, which are also the average $\EVb{G(s)}$. For each $W$, we compute 100 realizations of randomly chosen input and output weights by drawing $B,C$ as i.i.d. Gaussian variables with mean $\theta=\frac{1}{\sqrt{N}}$ and variance $\sigma^2=\frac{0.8}{N}$ (the scaling with $N$ allows comparison across different network size).  The resulting 100 realizations of $G(s)$ are plotted as blue traces. We see that of these realizations cluster around $\EVs{G(s)}$ more tightly as the network size increases. The gray areas are representing the 90\% confidence interval according to Eq. \eqref{eq:gaussian_CI_bound}. We emphasize that such convergence is strong in the sense it holds on a trial-to-trial basis for all frequencies $s=i\om$, as long as the network size $N$ is large.

\subsection{Parameters used in Fig. \ref{fig:mouse_brain}}
\label{sec:mouse_brain_area_list}
Here we use the connectivity between 213 cortical and thalamic areas in the mouse brain~\cite{oha2014}. The connection matrix in this dataset describes the density of axon projections from one area to another (Fig. \ref{fig:mouse_brain}A). We build a simple dynamic model by assuming that the node dynamics are identically determined by the exponential node filter $h_{\exp}(s)$ with a time constant of 100 ms, which is within the 50-350 ms range of intrinsic time constant used in the literature \cite{Murray:2014} ($\alpha=1/100$, Appendix \ref{sec:node_filter}). 

We consider the network transfer function $G(s)$ under two input patterns: thalamic input and cortical input. Here we include the list of areas used (by their abbreviations as defined in the mouse brain dataset~\cite{oha2014}). In the case of thalamic input, we send input uniformly to 11 sensory thalamic areas: AMd, AMv, LD, LGd, LP, MD, MGm, MGv, VAL, VPL, VPM. For cortical input, we send to 17 multimodal or associational cortical areas that are not primary or secondary sensory or motor areas: PTLp, FRP, PL, ILA, ORBl, ORBm, ORBvl, ACAd, ACAv, AId, AIp, AIv, RSPagl, RSPd, RSPv, TEa, PERI, ECT.

In Fig.~\ref{fig:mouse_brain}B, we plot the magnitude of chain motif cumulants $\abs{\gamma^n \kappa_n}$ against the order $n$. Here a constant $\gamma$ raised to proper power is inserted to set the scale for comparing motif cumulants across orders. The value of $\gamma$ is 90\% of the maximum value which satisfies the consistency requirement that $\kappa_n \gamma^n$ decays to 0 as $n\rightarrow \infty$. Note that these cumulants are computed by treating all nodes as belonging to a single population; a more complex but more accurate approach would be to consider subpopulation cumulants, as in Sec. \ref{S:multi_pop}.

In Fig. \ref{fig:mouse_brain}C, The $G(s)$ calculated based on original $W$ is plotted with red lines.  A sequence of blue lines depict successive (improving) approximations to this response computed by considering additional motif cumulants, that is keeping more terms of $\tilde{\kappa}_n$ in Eq. \eqref{E:define_mu_multi}. In the inset, the light blue curves are 100 samples produced by a node-degree preserving shuffle as explained in Fig.~\ref{fig:mouse_shuffle}. The effect is equivalent to setting every connection to a strength equal to the mean of the original log-normal weight distribution (red-dashed line), by keeping only the $\tilde{\kappa}_1$ term in Eq. \eqref{E:define_mu_multi}. A coupling strength is chosen at 90\% of the level of the maximum value that keeps the system stable.

\end{appendices}

\bibliographystyle{aip_title.bst}
\bibliography{papers3}

\end{document}